\documentclass[]{aastex631}

\usepackage{verbatim}
\usepackage{rotating}

\begin{document}

\title{SHORES: Serendipitous H-ATLAS-fields Observations of Radio Extragalactic Sources\\ with the ATCA. I: catalog generation and analysis}

\author[0000-0002-0375-8330]{Marcella Massardi}
\affiliation{INAF - Istituto di Radioastronomia - Italian ALMA Regional Centre,
Via Gobetti 101, 40129 Bologna, Italy}
\affiliation{Scuola Internazionale Superiore di Studi Avanzati, Via Bonomea 265, 34136 Trieste, Italy}

\author[0000-0002-6444-8547]{Meriem Behiri}
\affiliation{Scuola Internazionale Superiore di Studi Avanzati, Via Bonomea 265, 34136 Trieste, Italy}

\author[0000-0003-1394-7044]{Vincenzo Galluzzi}
\affiliation{INAF - Istituto di Radioastronomia, Via Gobetti 101, 40129 Bologna, Italy}
\affiliation{INAF-Osservatorio Astronomico di Trieste - Italian Astronomical Archives, via Tiepolo 11,34131 Trieste, Italy}

\author[0000-0002-1847-4496]{Marika Giulietti}
\affiliation{Scuola Internazionale Superiore di Studi Avanzati, Via Bonomea 265, 34136 Trieste, Italy}

\author[0000-0001-9808-0843]{Francesca Perrotta}
\affiliation{Scuola Internazionale Superiore di Studi Avanzati, Via Bonomea 265, 34136 Trieste, Italy}

\author[0000-0001-9680-7092]{Isabella Prandoni}
\affiliation{INAF - Istituto di Radioastronomia, Via Gobetti 101, 40129 Bologna, Italy}

\author[0000-0002-4882-1735]{Andrea Lapi}
\affiliation{Scuola Internazionale Superiore di Studi Avanzati, Via Bonomea 265, 34136 Trieste, Italy}
\affiliation{INAF - Istituto di Radioastronomia, Via Gobetti 101, 40129 Bologna, Italy}
\affiliation{Institute for Fundamental Physics of the Universe (IFPU), Via Beirut 2, 34014 Trieste}
\affiliation{Istituto Nazionale Fisica Nucleare (INFN), Sezione di Trieste, Via Valerio 2, 34127 Trieste, Italy}



\begin{abstract}
We introduce the Serendipitous H-ATLAS-fields Observations of Radio Extragalactic Sources (SHORES) multiple pencil beam survey that observed at $2.1$ GHz with the Australia Telescope Compact Array (ATCA) $29$ fields in total intensity and polarization within the \textit{Herschel}-ATLAS Southern Galactic Field. This paper presents the observations, calibration and analysis of the $27$ shallow fields that cover an overall area of $\sim26$ square degree with increasing sensitivity towards the phase centers of each pointing according to the ATCA $22$ m dish response function, down to $\sigma\lesssim 33\, \mu$Jy. Two additional (deep) fields have been observed to even higher sensitivity. All the SHORES observations have been calibrated to account also for linear polarization. Polarization and deeper field analysis will be presented in future papers. The SHORES shallow-field sample considered in the present paper counts $2294$ sources detected with \texttt{BLOBCAT} to signal-to-noise ratio $\rm{SNR}\gtrsim 4.5$. Simulations determined that our procedure and final catalog is $95\%$ reliable above $497.5\,\mu$Jy and $95\%$ complete to the $\rm{SNR}\gtrsim 4.5$ significance level. By exploiting ATCA E-W $6$ km configuration we reached resolutions of $3.2\times7.2$ arcsec, to which level $81\%$ of our sources are unresolved. We determined source counts down to the $150\,\mu$Jy level. For the sources with a counterpart in H-ATLAS, the FIR-radio correlation is calculated and discussed.
\end{abstract}



\section{Introduction} \label{sec:intro}

The flux density range below $1$ mJy at $\sim 2$ GHz has been for long an uncharted territory of the extragalactic source populations characterization \citep[e.g.][ Galluzzi et al. in prep.]{massardi10, dezotti10, mancuso17, simpson17, prandoni18, bonato21, behiri24}. 
While waiting for the Square Kilometer Array (SKA), several surveys have recently delivered their first results, improving the definition of the overall sub-mJy source-counts by observing large areas, comparable to their longstanding progenitor FIRST \citep{helfand15}, down to deeper sensitivity (VLASS \citealt{lacy20}; EMUPilot \citealt{norris21}; RACS-mid, \citealt{duchesne24}). 
Even deeper fields have been surveyed on smaller areas \citep[e.g.][]{gurkan22}, adding details to a picture that was already attempted by various deep-small-area surveys \citep[e.g.][]{gruppioni99, hopkins98, hopkins03, bondi03, prandoni00, prandoni18, butler18, heywood16, heywood22, damato22}. None of the above could avoid fighting, at different levels, against the combination of cosmic variance \citep{heywood13}, survey systematics introduced by calibration, deconvolution, source extraction algorithms, and bias corrections \citep[see ][; Galluzzi et al. in prep.]{condon12, heywood13} that introduce uncertainties in the counts definition. 

The overall picture that emerged demonstrated that a comprehensive characterization and classification of the source populations at sub-mJy flux density level does require the combination of multi-wavelength information: for this reason, the above-mentioned deep surveys mostly exploited cross-matches with X-ray, optical, and near-IR data.  

At sub-mJy flux density levels, the synchrotron emission powered by relativistic electrons generated by supernovae and their remnants starts to emerge over the AGN-dominated sources \citep{mancuso17}, with an integration time scale that depends on radiative losses and transport processes in the host galaxy. However, the presence of nuclear activity might enhance the synchrotron component with respect to the star formation-triggered one. Hence, accurate calibration of the far-infrared (FIR)/radio correlation (FIRRC) as a function of redshift and/or stellar mass \citep{novak17, delvecchio21} is essential to distinguish between radio emission driven by star formation and that powered by AGN activity. 

Some models for the cosmological evolution of radio emission from dusty star-forming galaxies (DSFGs) predict a decrease of the FIR/radio luminosity ratio with increasing redshift \citep{dezotti24}, as a consequence of the increase of cosmic ray energy losses at high z \citep{lacki10, schleicher13, schober16}, or because of the co-evolution of AGNs in dust enshrouded environments with active star formation in the early stages of galaxy evolution \citep[see ][]{lapi18}.

Recently, in order to characterize the FIRRC at high redshift, we \citep{giulietti22}
took advantage of the gravitational lensing magnification that characterizes the bright end of the FIR galaxy population and followed up with the Australia Telescope Compact Array (ATCA, project ID C3215, PI: Massardi) down to $\sigma\approx 0.06$ mJy per beam at $2.1$ GHz a complete sample of $30$ candidate lensed sources selected to be brighter than $100$ mJy at $500\,\mu$m in the Southern Galactic Pole (SGP) region of the \textit{Herschel}-ATLAS survey \citep{negrello17}.

The \textit{Herschel} Astrophysical Terahertz Large Area Survey \citep[H-ATLAS,][]{eales10} is the widest-area extragalactic survey undertaken via the \textit{Herschel} satellite with instruments surveying in the $100$, $160$, $250$, $350$ and $500\, \mu$m wavelength bands. The largest field of the survey is near the Southern Galactic Pole (SGP) and covers $270$ square degrees. It overlaps with multi-band surveys, including imaging and spectroscopic data in the optical bands provided by the \textit{Hubble Space Telescope} (HST), and near-IR imaging from the \textit{Spitzer Space Telescope} and \textit{Wide-field Infrared Survey Explorer} (WISE). In the radio band, the SGP region is covered by several surveys like SUMSS \citep{bock99}, NVSS \citep{condon98}, PMN \citep{griffith93}, AT20G \citep{murphy10, massardi11}, and, more recently, RACS \citep{mcconnell20}, and GLEAM-X \citep{hurley-walker22}, all targeting the $\gtrsim 1$ mJy population with a $\gtrsim 10$ arcsec resolution\footnote{RACS has the best sensitivity of $0.23$ mJy, but with a $15$ arcsec resolution in the range $887-1655$ MHz. The AT20G at the highest frequency of $20$ GHz reached a resolution of $2.4$ arcsec but only a sensitivity of $40$ mJy.} at the relative frequencies. 
Follow-up of many sources can be found in archives such as the ALMA Science Archive in the mm wavelengths regime and \textit{Chandra} in the X-ray \citep[e.g.][]{massardi16}.  

Characterizing the FIR-radio sky properties on scales larger than several arcsec also has relevance for cosmological studies since nuclear activity and star formation in DSFGs dominate the foreground emission and constitute the most critical contaminants in the component separation efforts in total intensity and polarization for present and future CMB facilities (e.g. Simons Observatory, LiteBIRD). Unfortunately, most of the deeper surveys listed above do not include polarization information on the target unresolved sources or have not released it yet. 
High fractions of polarization are considered hints of the presence of nuclear activity. \cite{galluzzi17} indicate a median polarization of $2.16\%$ for radio loud sources at $2.1$ GHz, unresolved on scales of $\sim10 $ arcsec  associated with synchrotron emission from jetted components. 
Stacking of \textit{Planck} observations \citep{bonavera17,trombetti18} with several arcmin resolution indicates that also DSFGs present a fraction of polarization that should not exceed $2.2\,\%$, associated with their synchrotron emission integrated over the line of sight and in the resolution element. 

For all of the above reasons, we decided to further deepen the data we had already in hand and perform new epochs of observations with the ATCA of the same H-ATLAS fields to improve $\mathcal{UV}$-coverage and sensitivity, serendipitously detecting sources in the fields that were initially targeted for other reasons. This resulted in a peculiar observing setup for the survey, and hence requiring caveats in the analysis that will be described in the following.
This is the first paper of a series that present the Serendipitous H-ATLAS-fields Observations of Radio Extragalactic Sources (SHORES) survey, performed in total intensity and polarization with the ATCA (project ID C3495) at $2.1$ GHz pointing to the $30$ candidate lensed galaxies in the H-ATLAS SGP surveys selected by \cite{negrello17}, with the main goal of combining radio and FIR information to characterize the sub-mJy $2$ GHz population.

Here we summarize the data collection and reduction procedure for the whole survey and the analysis for the 27 fields observed down to $\sim30\, \mu$Jy sensitivity level and for this reason dubbed as "shallow fields".
One of the fields, hereafter tagged as "deep field 1", has been observed more intensively to reach $8\, \mu$Jy. Two other fields partially overlap generating a mosaiced deeper region that was identified as our "deep field 2". The deep fields are being followed up also at $5.5-9.0$ GHz and $18-23$ GHz with ATCA (projects IDs C3502, CX542, C3605, some observations are still ongoing at the time of writing), and their analysis will be described in Paper II (Behiri et al. in prep.) and in future papers. All our observations include also polarimetric information, that will be described in Paper III (Galluzzi et al. in prep.).

The outline of the present paper is the following: in Section \ref{sec:observations} we present the target fields and the instrumental setup for all the epochs of ATCA observations in total intensity for the whole SHORES survey; in Sections \ref{sec:reduction} and \ref{sec:properties} we describe the data reduction, the extraction procedures and the properties of the survey, and in Section \ref{sec:catalog} we present the catalog for the $27$ SHORES shallow fields; in Sections \ref{sec:SourceCounts} and \ref{sec:FIRRC} we calculate source counts and, by matching with H-ATLAS counterparts, the FIRRC for our sources; finally in Section \ref{sec:conclusions} we summarize our findings. Appendix \ref{sec:lensed} includes a few notes on the lensed galaxies at the fields center.

Throughout this paper we define spectral indices $\alpha$ so that flux densities scale as $S_\nu\propto \nu^\alpha$, and we adopt a flat $\Lambda$CDM cosmology (\citealt{planck20}) with round parameter values $h\approx 0.67$, $\Omega_m\approx 0.3$ and $\Omega_{\Lambda}\approx 0.7$.

\section{ATCA Observations}\label{sec:observations}

This Section summarizes the observations collected to build up the SHORES survey, including information on target fields and observing epochs.


\begin{figure}[ht!]
\plotone{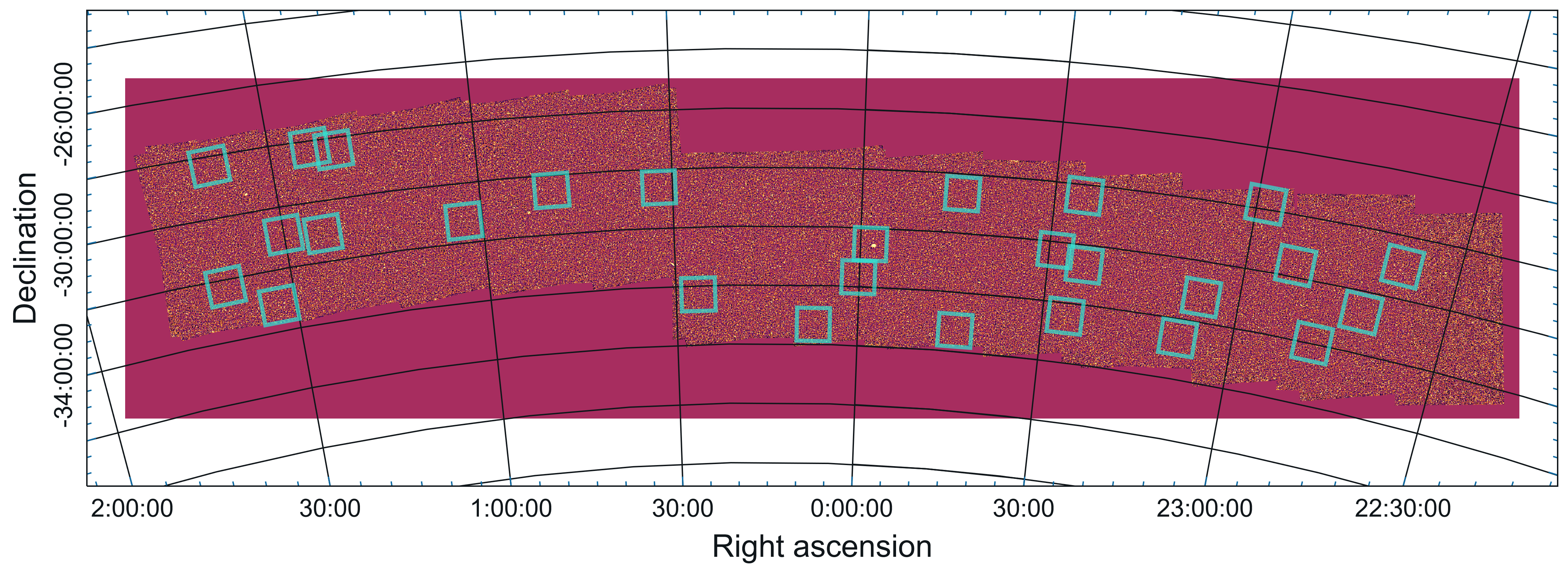}
\caption{H-ATLAS map of the SGP region with superimposed the footprints of the SHORES shallow fields. The center of each field coincides with one of the candidate lensed galaxies in \cite{negrello17}.}
\label{fig:map}
\end{figure}

Gravitational lensing magnifies the flux densities of the lensed sources: DSFGs that dominate the source counts in the FIR domain are thus pushed to the counts' brightest end if affected by lensing. With this criterion, \cite{negrello17} identified a sample of candidate lensed galaxies by selecting source brighter than 100 mJy at $500\, \mu$m in the fields of the H-ATLAS survey. Thirty of them were in the Southern Galactic Pole (SGP) region with declination $\delta\sim -30^\circ$ and right ascension $22\lesssim$ RA $\lesssim 2$ hours (see map in Figure \ref{fig:map}). 

The sample was also targeted with optical observations \citep[VIKING, HST, see ][ and references therein]{borsato23}, and the region is also partially covered by Spitzer \footnote{\url{https://irsa.ipac.caltech.edu/Missions/spitzer.html}} and wholly by WISE \footnote{\url{https://wise2.ipac.caltech.edu/docs/release/allsky/}}. Several of the lensed sources have been targeted with ALMA millimetric band follow-ups to confirm their lensed nature, and with \textit{Chandra} or \textit{XMM-Newton} to identify the presence of nuclear activity. 

In 2017-2019 we performed an observing campaign of about 11 hours of overall observing time targeting the SGP candidate lensed sources by \cite{negrello17} to investigate their FIR-radio properties. Details on the observations, data reduction and results can be found in \cite{giulietti22}. Data reached 0.06 mJy beam$^{-1}$ sensitivity (1$\sigma$), with resolution $\sim 10$ arcsec, thanks to the East-West 6-km extended configuration of the ATCA array. Reduction and imaging focused on detecting and analyzing the target that was in the phase center of each field. Each candidate lensed galaxy was targeted separately, despite two of such galaxies (namely HATLASJ005132.8-01848 and HATLASJ005132.0-302011) happen to be closer than the 22.2 arcmin FWHM  of the 2.1 GHz ATCA FoV, and in principle observable with a single pointing.
Then, $11$ targets in the H-ATLAS SGP were detected above 5$\sigma$, even if the resolution was not sufficient to confirm them as lensed sources.
However, the images immediately presented a large number of H-ATLAS sources detected below 1\ mJy within each ATCA field of view (FOV). This encouraged us to embark in a deeper survey, the SHORES `multiple pencil-beams survey'.
We targeted the same fields at $2.1$ GHz with the 2 overlapping $2\ $GHz-wide sidebands of the CABB correlator (in its standard 1 MHz of channel size configuration) in several epochs between October 2022 and July 2024.

After cross-matching our targets with existing surveys, one of the fields (s0009-3018\footnote{Hereafter we tagged our observed fields according to the coordinates of the central HATLAS source with the nomenclature ``s\emph{hhmm-ddmm}''}, centered on HATLASJ000912.7-300807), emerged as not hosting bright ($\gtrsim 10$ mJy) sources in radio catalogs, while being (at least partially) covered by \textit{Spitzer}, MUSE, and HST. Therefore, we selected it to be observed to even deeper sensitivity with additional dedicated epochs at $2.1$ GHz. This field is identified as ``SHORES deep field 1''.
Also in the most recent epochs, the two overlapping pointings on the two close targets were observed as separated targets, resulting in an over-sampled mosaic at the imaging stage identified as ``SHORES deep field 2''.
The deep fields have been observed together with all the other fields at $2.1$ GHz and in additional dedicated epochs, and followed up also at $5.5-9.0$, and $18-23$ GHz. Their observations, multi-frequency catalog and population properties will be detailed in a separate paper. Dedicated follow-up on some peculiar targets with higher frequencies or compact configurations are still on-going and will be further described in dedicated papers.

As in previous observations, we exploited the most extended East-West 6-km configuration, including antenna CA06, to maximize resolution and minimize blending effects. Additional epochs were allocated as green time while the requested configuration was available: they helped us to better cope with the telescope RFI and improve $\mathcal{UV}$-plane coverage to favor imaging. A total of $143,5$ hrs of observing time have been, to date, allocated for $2.1$ GHz observations and $86$ hr for follow-ups of the deep fields at other frequencies. A summary of the overall SHORES survey observing epochs for the various project IDs, telescope setup, and observed targets is presented in Table \ref{tab:epochs}. 

In this work, we focus on the $2.1$ GHz data reduction and imaging procedures that were applied to all the fields and on the catalog extracted in total intensity from all the 27 ``SHORES shallow fields''. They were observed at $2.1$ GHz in sequences of $3$ minutes cuts for the whole available time with interleaves for phase-leakage calibrator observations every $\sim 30$ minutes. Weather and telescope response were excellent through almost all the observing epochs. 

\begin{table*}[ht]
    \centering
    \begin{tabular}{c|c|c|c|c|c}
    \hline
       ATCA& Date & UT time &  Frequency & Targets &ATCA     \\
       Project id & &   range &  [GHz]   &  &configuration \\
    \hline
        C3215&2017Nov22& 14.00-17.30& 2.1& SF     & 6D   \\
        C3215&2017Dec15& 03.00-15.00& 2.1& SF     & 6D   \\
        C3215&2019Jul31& 16.30-20.30& 2.1& SF     & 750C
        \\
        C3502&2022Oct09& 04.30-18.30& 5.5-9.0& DF1& 6D   \\
        C3495&2022Nov29& 04.00-16.30& 2.1& SF     & 6D   \\
        C3495&2022Nov30& 04.30-16.30& 2.1& SF     & 6D   \\
        C3495&2022Dec16& 05.00-09.00& 2.1& SF     & 6D   \\
        C3495&2022Dec20& 07.30-13.00& 2.1& SF     & 6D   \\
        C3502&2022Dec23& 05.30-14.00& 5.5-9.0& DF1& 6D   \\
        C3495&2022Dec24& 01.00-14.00& 2.1& DF1+someSF  & 6D
        \\
        C3495&2022Dec25& 09.30-14.00& 2.1& DF1    & 6D   \\
        C3495&2022Dec24& 01.30-14.00& 2.1& DF1+someSF  & 6D
        \\
        C3502&2022Dec28& 06.00-14.00& 5.5-9.0& DF1& 6D   \\
        C3495&2022Dec29& 01.00-14.00& 2.1& DF1+someSF  & 6D
        \\
        C3495&2022Dec30& 01.30-14.00& 2.1& SF     & 6D   \\
        C3502&2023Jan14& 00.30-13.00& 16.7-21.2& DF1  & 6D
        \\
        C3495&2023Jan25& 00.00-10.00& 2.1& DF1    & 6D   \\
        CX542&2023Dec25& 09.00-16.00& 2.1& DF2    & 6D   \\
        CX542&2024Feb29& 22:00-09.00& 5.5-9.0 & DF1 & 6D \\
        CX542&2024Mar14& 21.30-08:30 & 5.5-9.0& DF1 & 6D  \\
        CX542&2024Mar18& 01:30-09:30 & 2.1 & DF2 & 6D   \\
        C3605&2024Apr28& 18:30-05:30 & 5.5-9.0& DF2 & 6D \\
        C3605&2024May03& 18:00-04:00 & 5.5-9.0& DF2 & 6D \\
        C3605&2024Jun09& 15:32-03:30 & 5.5-9.0& DF2 & 6D
        \\
        C3605&2024Jul26& 18:30-04:00 & 2.1& SF & H214   \\
        \hline
        \end{tabular}
    \caption{Summary of SHORES observation epochs with the ATCA. `DF' indicates epochs dedicated to the deep field, while `SF' indicates observations dedicated to all the survey fields (i.e. including both shallow and deep fields).}
    \label{tab:epochs}
\end{table*}

\section{Data reduction}\label{sec:reduction}

In this Section details on the data reduction procedure, including editing, imaging, and (self-)calibration are presented.

\subsection{Editing and calibration}

Data editing and calibration were performed using the \texttt{Miriad} \citep{miriad} suite. 
At 2.1 GHz, for each epoch only the best of the two $2$ GHz-wide overlapping sidebands was preserved and flagged to remove RFI and misbehavior, by combining manual (using \texttt{UVFLAG} and \texttt{BLFLAG} tasks) and automatic (\texttt{PGFLAG}) approaches in an iterative sequence. In the end, about $30-50\%$ of each epoch data were flagged and discarded from any following analysis.

Given that the standard bandpass and flux density (primary) calibrator quasar PKS1934-638 (a well-known GigaHertz Peaked Spectrum, GPS, source) was observable in all the slots, we followed the standard recipe for cm-data reduction described in the ATCA Users' Guide\footnote{\url{https://www.narrabri.atnf.csiro.au/observing/users_guide/html/atug.html}.}. 

As phase calibrators at $2.1\,$GHz we used PKS2255-282 (Seyfert 1 at $z\approx 0.926$ as bright as $\sim 2.15$ Jy in total intensity) and PKS0008-421 (QSO at $z\approx 1.12$ with $\sim 3.06$ Jy total intensity flux density) for fields with $22\lesssim$ RA $\lesssim 24$ hr and $00\lesssim $ RA $\lesssim 02$ hr, respectively. Gain solutions were determined by using the \texttt{Miriad} task \texttt{GPCAL} independently on $4\times512$ MHz-wide sub-bands in which we split the observed CABB bands (namely centered at $1332$, $1844$, $2356$, and $2868$ MHz). However, in some epochs, where the lowest (centered at $1332$ MHz) and sometimes also the highest (centered at $2868$ MHz) sub-bands turned out to be heavily affected by RFI (e.g. as in the case of epochs 2022Nov29, 2022Nov30, and 2023Dec25) only two $1024\,$MHz-wide sub-bands were considered, as this guaranteed enough signal to noise for deriving corrections.

Moreover, since PKS1934-638 can be assumed to be unpolarized in the $16$ cm frequency-band (since its linear polarization fraction is lower than what is achievable with a first-order leakage term correction, i.e. $\lesssim0.1-0.2\%$), according to the standard calibrator recipe we determine instrumental polarisation leakages (the so-called ``D-terms'') from an additional bright polarized object. 
Typically we exploited the phase calibrator as leakage calibrator too: the BLLac PKS0008-421 (at $z\approx 1.12$, hence compact for ATCA resolution at this frequency) is $\sim 1.5\%$ polarized at $2.1$ GHz. To check leakage stability, when possible we also included the radio galaxy PKS2341-244 (at $z\approx 0.59$), and the BLLac PKS0426-380 ($z\approx 1.11$), known to be polarized at a few percent level\footnote{PKS2341-244 is $\sim 3.1\%$ polarized at $1.4$ GHz \citep{farnes14}, and PKS0426-380 is reported about $\sim 6\%$ polarized at $1.4$ GHz by \citet{lamee16}, while \citep{massardi11} reports $3.1\%$, $1.6\%$ and $1.4\%$ at $4.8$ GHz, $8.6$ GHz and $18$ GHz respectively.}. 
Leakage calibrator observations included at least $4-6$ observing cuts at different parallactic angles (coverign more than $90$ deg) to both determine the intrinsic Stokes Q and U for each source, and refine leakage D-terms. 
Also, similarly to the gain calibration, leakage solutions were determined independently from each of the $4\times512$ MHz-wide sub-bands in which we split the data.
Further details on the leakage calibrations performed in the various epochs and results of the polarization data analysis will be collected in a paper dedicated to the polarization properties of the SHORES sources.

\subsection{Imaging and self-calibration}

\begin{figure}
   \centering
   \includegraphics[width=0.5\textwidth]{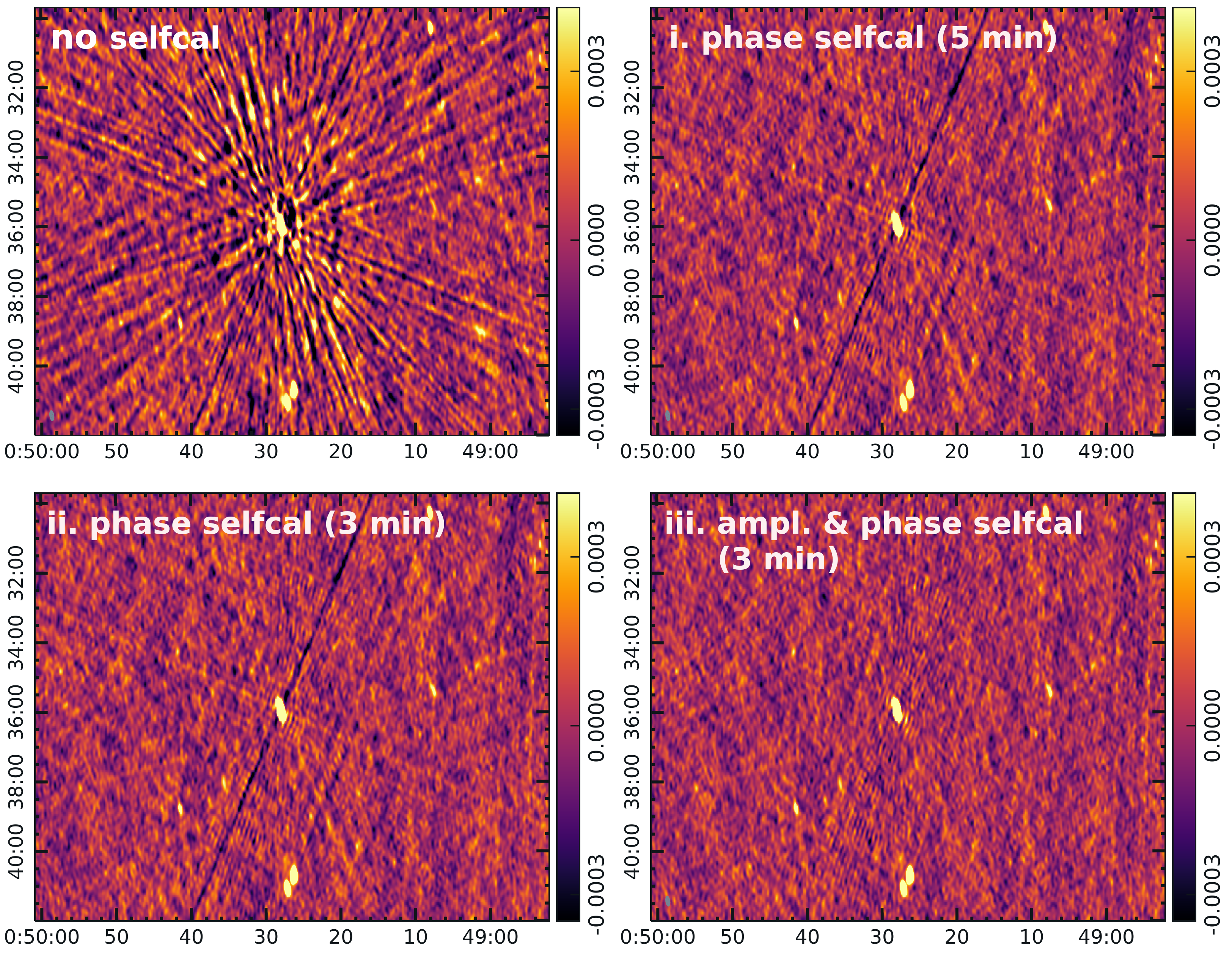}
    \caption{Comparison of Stokes' I maps at $1332\,$MHz around the brightest source ({\it i.e.} $36\,$mJy) of the field s0048-3031. Top panels: {\it (Left)} initial map without self-calibration (noise rms measured at the border of the FOV with no sources is $~107\,\mu$Jy/beam); {\it (Right)} the same region after a first round of phase self-calibration ($5\,$min solution interval, rms $~82\,\mu$Jy/beam). Bottom panels: {\it (Left)} after the second round of phase self-calibration ($3\,$min solution interval, rms $~80\,\mu$Jy/beam); {\it (Right)} after the third and final round of self-calibration, {\it i.e.} both in amplitude and phase ($3\,$min solution interval, rms $~77\,\mu$Jy/beam).} 
  \label{fig:selfcalexs0048}
\end{figure}

After the main calibration has been performed for each epoch exploiting an external calibrator as phase-reference, we considered the possibility to refine gain solutions via self-calibration. To this purpose, we applied all the calibration Tables for each target/field at each epochs (with \texttt{Miriad} task \texttt{UVAVER}), and concatenated the visibilities from the different epochs (with \texttt{Miriad} task \texttt{UVCAT}). 
Then, we inspected again visibilities for each baseline in order to flag any residual RFI that may hamper imaging and self-calibration attempts. 

We performed a preliminary imaging with \texttt{Miriad} (exploiting the tasks \texttt{INVERT} and \texttt{MFCLEAN} for inverse Fourier Transform and deconvolution) in order to identify those fields and sub-bands where at least a source with a surface brightness feature brighter than $15$ mJy/beam was present within the FOV. In fact, taking into account data flagging and other observing conditions during our campaign, this value allows to reach a $\rm{SNR}$ larger than $10$ on a given $512$ MHz-wide sub-band over a time span of $3$ min, which is suitable for determining gains at $2.1$ GHz with the ATCA in the used extended E-W configuration. We also tried shorter solution intervals (e.g., $2$ and $1$ min) in case of relatively bright sources at a few hundreds of mJy level, and relaxing the brightness threshold, finding no significant improvements either in the gain solutions, or in the derived imaging. Thus, in cases where the source brightness is $\gtrsim 15$ mJy/beam, for at least a source in the field, we performed three loops of imaging and self-calibration (by exploiting the \texttt{Miriad} task \texttt{GPSCAL}), the first two in phase only, and the last in amplitude and phase. Solution intervals adopted were $5$, $3$ and $3$ min, respectively. Given the wider (up to a factor $~2$) FOV at the lowest frequency sub-band (i.e., $1332$ MHz) compared to the higher sub-bands, as well as the typically steep spectrum for radio sources, it was expected that more bright sources were present at the lowest sub-band. In fact, in $22$ over $27$ fields the self-calibration was possible at $1332$ MHz, in $11$ it was possible at $1844$ MHz, compared to only $5$ and $2$ fields at $2356$ and $2868$ MHz, respectively. 

As an example of effectiveness of the applied self-calibration procedures, Figure~\ref{fig:selfcalexs0048} displays the cutout from the Stokes' I map at $1332$ MHz of about $10$ arcmin around the brightest (and compact) source found in the field s0048-3031, with a flux density of $36$ mJy: in the upper-left panel with no self-calibration dominant phase-related antisymmetric image artifacts are visible, and the root mean squared (rms) of pixels values estimated from a region in the outskirt of the FOV with no detected sources is about $107\,\mu$Jy; the upper-right and the bottom-left panels shows the outcome of a first, and a second round (respectively) of phase self-calibration, where only symmetric (amplitude-errors related) features are present. The resulting rms levels are close, being $\sim 82\,\mu$Jy and $\sim 80\,\mu$Jy, respectively. Then, we attempted a third round, this time both in amplitude and phase by keeping the solution interval of $3$ min. Considering that the source was not so bright and that number of ATCA antennas is small, the final map shows an appreciable reduction of residual artifacts and a further reduction of the rms to $77\, \mu$Jy/beam. In case of brighter sources, we could reduce the rms by even larger factors (up to 2.3).      
We performed self-calibration loops for all the fields and sub-bands where this was possible. For fields with less than 4 self-calibrated sub-band (i.e., at least $1332$ MHz data) we transferred the derived gain tables to the other sub-bands.

For fields s2242-3241 and s2358-3232 no self-calibration was possible due to lack of bright sources, but they are close (within $3-4$ deg) to bright sources (at a few mJy level in all the sub-bands) in nearby fields s2237-3058 and s2344-3039, respectively. Therefore, we copied the gain tables from the latter. 

The final (`full-Stokes') imaging was performed by exploiting \texttt{WSCLEAN} \citep[v. 3.4,][]{offringa14}: for this reasons all the visibility files were converted to MeasurementSet (as \texttt{WSCLEAN} does not support \texttt{Miriad} visibility format). This choice was motivated by two reasons: \texttt{WSCLEAN} is an efficient wide-field imager adopting a $\mathcal{W}$-stacking algorithm (it applies $\mathcal{W}$-correction after the inverse Fourier transform of gridded layers at different values of $\mathcal{W}$, which describes the curvature of $\mathcal{UV}$-space) and offers the possibility to perform a joint deconvolution of the different Stokes parameters (so to consider the polarization information across the Stokes). 

To maximize the surveyed area and the number of detected sources we attempted to image as far from the phase center as still acceptable for the ATCA beam shape. However, this requires to characterize the Primary Beam (PB-)correction for ATCA out to large  distances from the phase center, as described in Sect. \ref{sec:pbcorr}. Therefore, we produced only non PB-corrected images, applying the correction as a-posteriori step at the catalog level, after performing the source detection on non-PB corrected maps.

Finally, for each field, we imaged over a region $\sim 3$ times the nominal FOV-FWHM (i.e., $3\times 22.2$ arcmin) at $2.1$ GHz, over which we could recover a suitable PB characterization, hence each map covers $1.116 \times 1.116$ deg$^2$ with a pixel resolution of $1$ arcsec (in order to sample the synthesized beam by at least $3\times 5$ pixels). We adopted a robust (Briggs parameter set to $0.5$) weighting scheme in order to strike a balance between artifacts suppression and the best sensitivity. We exploited the auto-masking feature to identify all emissions above $5\sigma$ significance level, and performed a deep cleaning down to the noise level ($1\sigma$, auto-threshold set to $1$) over those regions. The resulting synthesized beam is about $3.2\times 7.2$ arcsec for each field. In the most used configuration (6D) the shortest baseline is $\sim337$m corresponding to $\sim85$ arcsec.

\section{Survey properties}\label{sec:properties}

In this Section we describe the procedures for source extraction, the computation of the primary beam correction, and the assessment of the completeness and reliability of the survey via dedicated simulations.

\subsection{Source detections}

\begin{table}
    \centering
    \begin{tabular}{cccc}
    \hline
    SNR &  $\geq3$ & $\geq4.5$ & $\geq5$ \\
    \hline
    \texttt{BLOBCAT}   & 13662 & 2697 & 2008\\
    \texttt{AEGEAN}    &  2867 & 2768 & 2615 \\
    \texttt{PySE}      &  8015 & 4792 & 4310\\
    \hline
    \texttt{BLOBCAT}  and \texttt{AEGEAN}  & 2197 & 2122& 1937\\
    \texttt{BLOBCAT}  and \texttt{PySE}    & 3917 & 1210& 1051\\
    \texttt{AEGEAN}   and \texttt{PySE}    & 6345 & 2115&  1690\\
    \hline
    \end{tabular}
    \caption{Number of detections with varying tested methods and significance levels.}\label{tab:detections}
    \label{tab:my_label}
\end{table}

\begin{figure}
    \centering
    \includegraphics[width=0.45\linewidth]{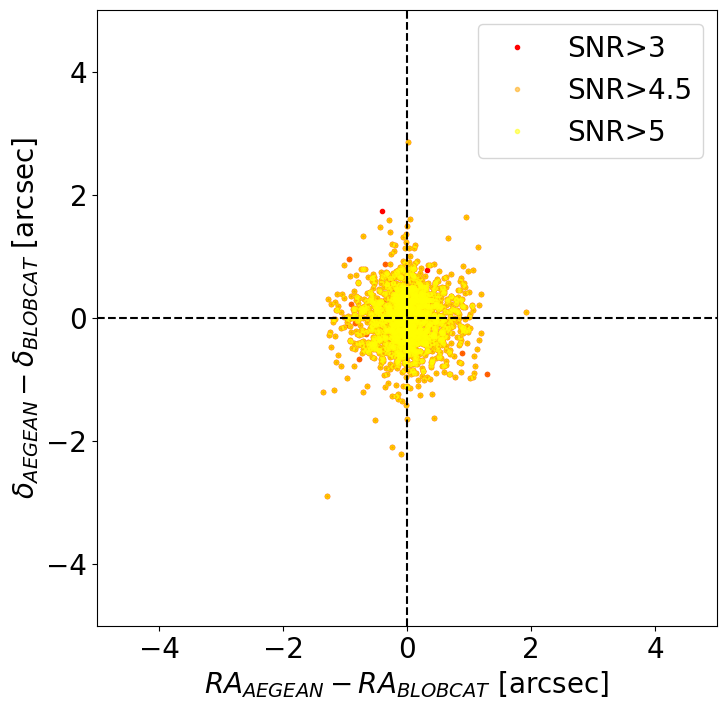}
    \includegraphics[width=0.45\linewidth]{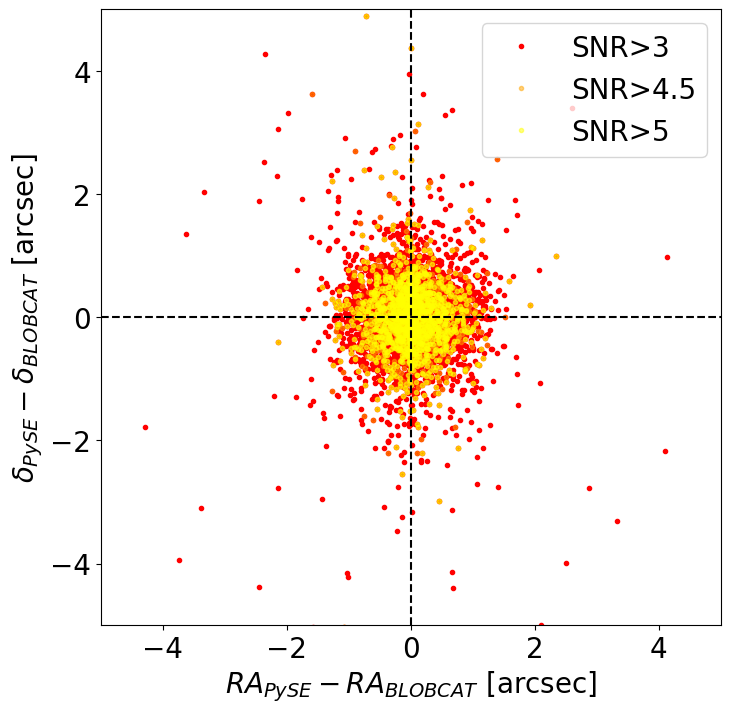}
    \caption{Positional discrepancy between detections. Colours refer to the \texttt{BLOBCAT} SNR.}
    \label{fig:dcoord}
\end{figure}

\begin{figure}
    \centering
    \includegraphics[width=0.30\linewidth]{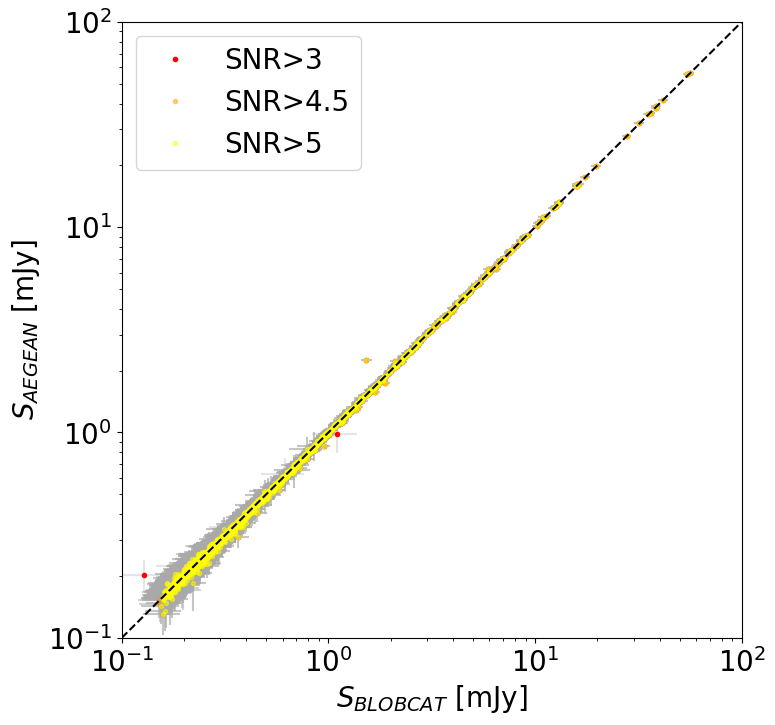}
    \includegraphics[width=0.30\linewidth]{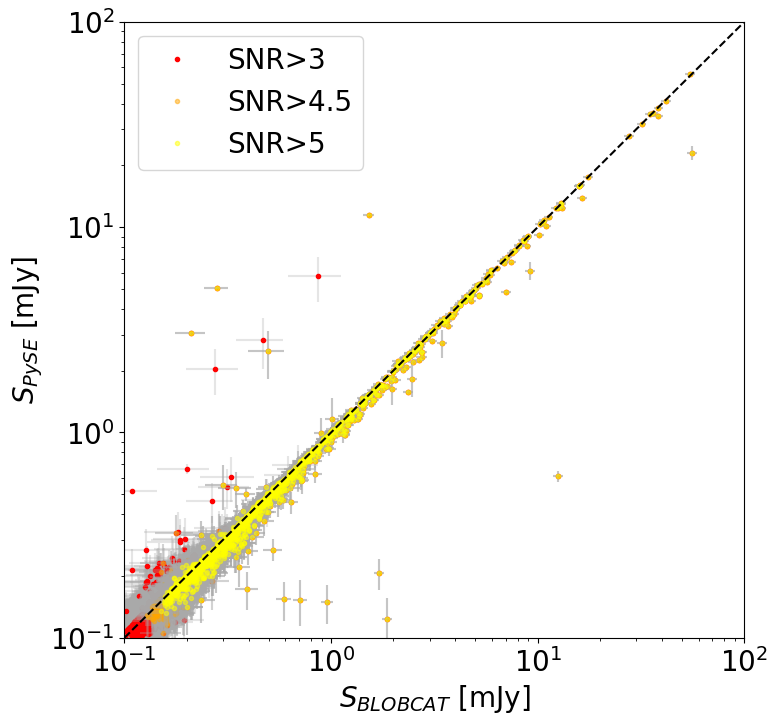}
    \includegraphics[width=0.30\linewidth]{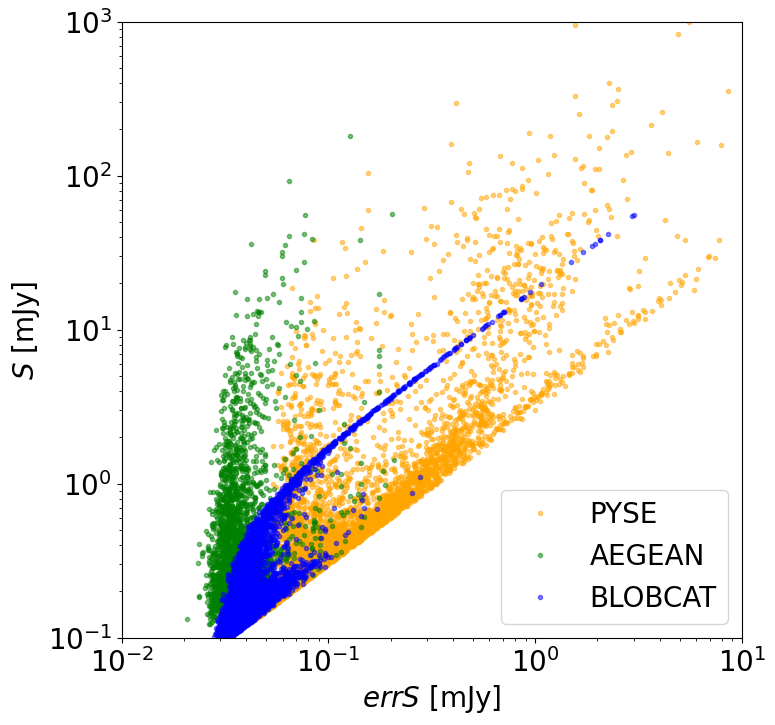}
    \caption{\textit{Left and central panels} Comparison of flux densities detected with the various methods. Colours refer to the \texttt{BLOBCAT} SNR. \textit{Right panel} Comparison of peak flux densities and their errors as extracted by each method color-coded as \texttt{BLOBCAT} in blue, \texttt{AEGEAN} in green, \texttt{PySE} in orange. }
    \label{fig:dfluxes}
\end{figure}

Several methods for source detections are nowadays available in the literature, that differently adapt to survey goals, noise patterns, data format and origin. 
We stress that a comprehensive comparison of the various methods is well beyond the scope of the present paper. However, to pursue suitable and well-understood performances for the whole procedure adopted to generate the SHORES catalogs, we decided to apply more than one method. Following the indications of \cite{hopkins15}, we choose some among the methods that are indicated to give better results in terms of reliability, completeness and flux density recovery for interferometric data, and that have already been used on ATCA data \citep[see also][]{boyce23}. The following three methods of source detection have been applied to the non-primary-beam-corrected maps (i.e. the whole band, hereafter indicated as multifrequency-synthesis, ``mfs", map at 2.1\,GHz, and each of the four 512\,MHz-wide maps of the sub-bands) of each SHORES field.

\begin{description}
    \item[\texttt{AEGEAN} \citep{hancock12,aegean}] \texttt{AEGEAN} uses a flood fill algorithm to isolate individual islands of pixels from within a signal to noise ($\rm{SNR}$) map. It assumes a compact source structure to fit the multiple components in one island and each of them is characterised by at least one Gaussian component. \texttt{AEGEAN} inputs are the target map and the noise and background maps produced by \texttt{BANE}, its inbuilt background and noise calculation algorithm. \texttt{BANE} calculates the mean and the standard deviation on a sparse grid of pixels and then interpolates them to give the final background and noise images and, to avoid contamination from source pixels, it performs sigma clipping. The tool provides a size of the major and minor axis and position angle of the Gaussian component that better fits the island.

    \item[\texttt{PySE} \citep{pyse}] \texttt{PYSE}, initially developed for LOFAR images, performs a $k\sigma$-clipping around the median of each box in which it divides the map. Each box's calculated modes and standard deviations are then interpolated to produce the background and noise maps. In this way it is particularly sensitive to extended noise patterns as those generated by very extended and bright sources. According to \cite{hopkins15} it performes reasonably well above $4\sigma$ levels.  
    
    \item[\texttt{BLOBCAT} \citep{blobcat}] \texttt{BLOBCAT} also uses a flood fill algorithm to isolate individual islands of pixels from within a signal to noise ($\rm{SNR}$) map. The $\rm{SNR}$ map is formed by taking the pixel-by-pixel ratio between the input map and the noise image generated with other tools. The software is designed to process radio-wavelength images of both Stokes I intensity and linear polarization, the latter formed through the quadrature sum of Stokes Q and U intensities or as a by-product of rotation measure synthesis. Flux densities are then corrected for peak and clean bias and for bandwidth smearing if a map, or a constant value for it, is provided as input (see discussion below). Among the outputs the tool provides an estimate of the blob size in units of the Gaussian size that a point source with the retrieved peak flux would have: this is a measure of the source size.     
 \end{description}

For \texttt{BLOBCAT} in the following we report results obtained with the \texttt{PYSE}-generated noise map, but we tested also the \texttt{BANE}-generated maps with indistinguishable outcomes.
If not differently mentioned, we hereafter define as `SNR' the ratio between the flux density of the detection and the local noise level in the noise maps. We cross matched candidate sources detected with different methods with $\rm{SNR}\geq3$ associating detections that fall within a radius equal to the maximum between our resolution and the candidate sources sizes. Table \ref{tab:detections} summarizes the numbers of $\rm{SNR}\geq3$ detections matched with the various methods at various significance level. For the detections that appear in common to more than one method we verified that the positional discrepancy reduces with the SNR and the flux densities agree among the methods (see Figures \ref{fig:dcoord} and \ref{fig:dfluxes}).  

BLOBCAT detected 13688 candidate sources with signal-to-noise ratio $\rm{SNR}\geq3$ of which 2697 have $\rm{SNR}\geq4.5$. 
AEGEAN detected 2867 candidate sources with $\rm{SNR}\geq3$, of which 2768 have $\rm{SNR}\geq4.5$. 
Among the 8015 detections with $\rm{SNR}\geq3$ in PySE, 4792 have $\rm{SNR}\geq4.5$.
This seems to indicate that both PySE and BLOBCAT may be more affected by spurious detections than AEGEAN at low significance levels. About $96.5\%$ of the \texttt{BLOBCAT} SNR$>5$ detections are detected also by \texttt{AEGEAN} at the same significance, while only $52\%$ are found by \texttt{PySE} at the same significance. The fractions reduce to the $78.7\%$ and $44.8\%$ respectively at $\rm{SNR}\geq4.5$. Only $74.1\%$ of the \texttt{AEGEAN} detections at high significance level are confirmed also with \texttt{BLOBCAT}. This indicates a difference in the definition of the significance level, that appears a bit more optimistic in the case of \texttt{AEGEAN}, in particular in presence of bright sources. Furthermore, there is a difference among the two methods in the definition of flux density errors. In \texttt{BLOBCAT}, the flux density and flux density  error include corrections for peak bias, clean bias, bandwidth smearing for which the terms proportional to the source flux density dominate with respect to the background noise. Such corrections are not included in \texttt{AEGEAN} (see for comparison Figure \ref{fig:dfluxes}). A visual inspection of all the detections confirmed that the noise values indicated by \texttt{BLOBCAT} for each detection accounts for the noise pattern surrounding each source and provides a realistic estimate of what could be measured around the sources, reducing the rate of spurious high significance detections in the surroundings of very bright sources. The behavior and the measured noise values get comparable among the two methods as the source flux density gets lower.

In summary, \texttt{AEGEAN} and \texttt{BLOBCAT} perform similarly on our maps, with the discussed differences in the noise definition, while \texttt{PySE} presents more discrepancies from the other methods, in particular at the faintest fluxes. The conservative definition of the flux density error of \texttt{BLOBCAT}, its applied corrections, and the joint detection in I, Q and U Stokes maps perfectly fits the purposes and design of SHORES and for these reasons we selected it as our main tool, exploiting the other methods as independent confirmations, when necessary. Therefore in the following we will report results obtained with the \texttt{BLOBCAT} flux densities, but we nevertheless generated a catalog of all the detections common to at least two methods, as described below. 

\subsection{Primary beam correction}\label{sec:pbcorr}

\begin{table}
    \centering
    \begin{tabular}{c|c|c|c|c|c|c}
    \hline
         Annulus& Distance   &A$_{eff}$ & Median            & Median            & $\#$ of detected  & Minimum \\
                & [arcmin]   &[deg$^2$] & $S_{pix}$ [mJy]   & $\sigma_{pb}$[\%] &  sources          & $S_{SNR>4.5}$[mJy]      \\
    \hline
         0      &  11.1      & 2.49     & 0.06              &  3                & 715               & 0.16  \\
         1      &  11.1-22.2 & 7.31     & 0.3               &  8                & 946               & 0.29    \\
         2      &  22.2-33.3 & 9.25     & 0.8               & 13                & 573               & 2.05    \\
    \hline
    \end{tabular}
    \caption{Summary of the properties of the annuli due to the telescope response function and identified on the field maps, including the radial distance from the pointing center (see \S \ref{sec:pbcorr}), the effective area, the median of the pixel distribution of the annulus over all the SHORES shallow fields (see \S \ref{sec:area}), the median percentage correction of detected flux densities applied to the flux density errors to account for calibration uncertainties (see \S \ref{sec:noise}), the overall number of sources detected with \texttt{BLOBCAT} above $4.5\sigma$ and listed in the SHORES catalog, and the minimum of their primary beam corrected flux densities (see \S \ref{sec:catalog}).}
    \label{tab:annuli}
\end{table}

\begin{table}
    \centering
    \begin{tabular}{cccccc}
    \hline
      n & 1.332 GHz & 1.744 GHz & 2.356 GHz &2.868 GHz & 2.1 GHz mfs  \\
      \hline
0  & 9.9991E-01  &  9.9998E-01  &  1.0007E+00  &  1.0001E+00  &  1.0001E+00  \\
1  & 4.8744E-03  &  -3.2138E-03  &  3.5764E-03  &  -4.7504E-03  &  1.0650E-02  \\
2  & -3.9644E-03  &  -2.9066E-03  &  -5.0351E-03  &  -4.8295E-03  &  -6.5948E-03  \\
3  & 3.1269E-04  &  -1.7043E-05  &  -3.5922E-04  &  -8.9423E-04  &  1.3416E-05  \\
4  & -2.7387E-05  &  3.2981E-06  &  6.2998E-05  &  1.4505E-04  &  2.9828E-05  \\
5  & 1.5442E-06  &  1.9846E-07  &  -3.0334E-06  &  -7.8255E-06  &  -1.6489E-06  \\
6  & -4.6575E-08  &  -1.3757E-08  &  6.9728E-08  &  2.0536E-07  &  3.9516E-08  \\
7  & 7.0896E-10  &  2.7044E-10  &  -7.9248E-10  &  -2.6703E-09  &  -4.5427E-10  \\
8  & -4.3094E-12  &  -1.8049E-12  &  3.5882E-12  &  1.3818E-11  &  2.0369E-12  \\
\hline
    \end{tabular}
    \caption{Parameters of the polynomial expansion for the SHORES measured primary beam correction.}
    \label{tab:pbcorr}
\end{table}

\begin{figure}
    \centering
    \includegraphics[width=0.45\textwidth]{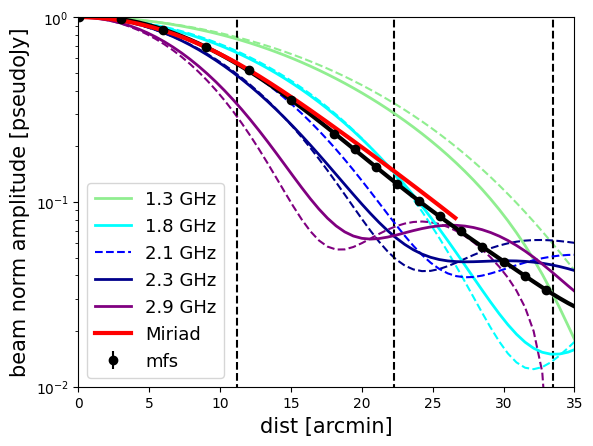}
    \includegraphics[width=0.45\textwidth]{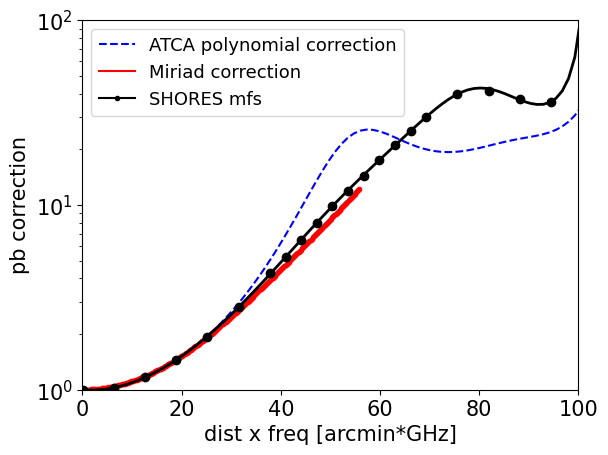}
    \caption{\textit{(Left panel)} Profile of the ATCA primary beam response function that we measured on PKSJ0537-441 in the mfs image (black line and dots) and for each of our sub-bands (solid lines) compared with the multi-frequency synthesis profile implemented in \texttt{Miriad} (red line), and reported in ATCA polynomial expansion for each of the listed sub-bands (dashed lines). The dashed black dashed vertical lines divide the plots into the three annuli discussed in the text. \textit{(Right panel)} Primary beam correction as calculated from the beam shapes of the left panel for our mfs data (black line and dots) compared with \texttt{Miriad} mfs (red line) and $2.1\ $GHz polynomial expansion corrections (blue dashed line).} 
    \label{fig:pbcorr}
\end{figure}

\begin{figure}
    \centering
    \includegraphics[width=0.30\linewidth]{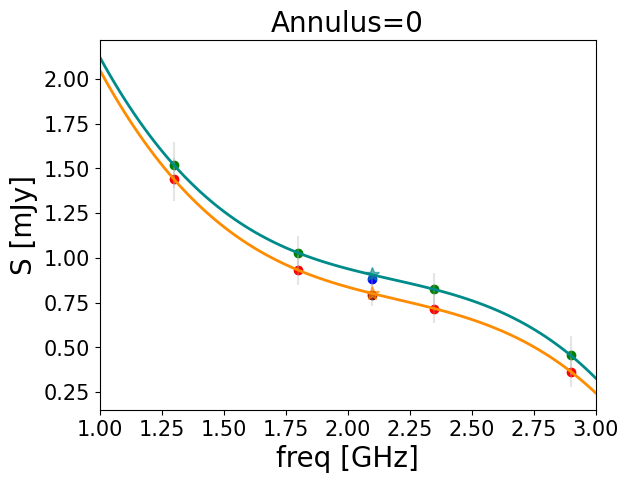}
    \includegraphics[width=0.30\linewidth]{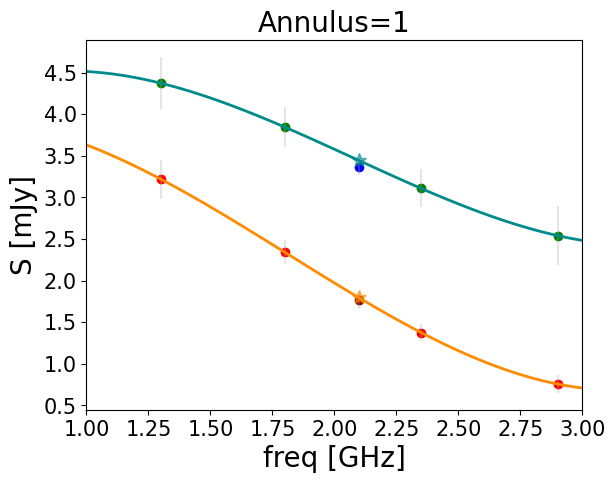}
    \includegraphics[width=0.30\linewidth]{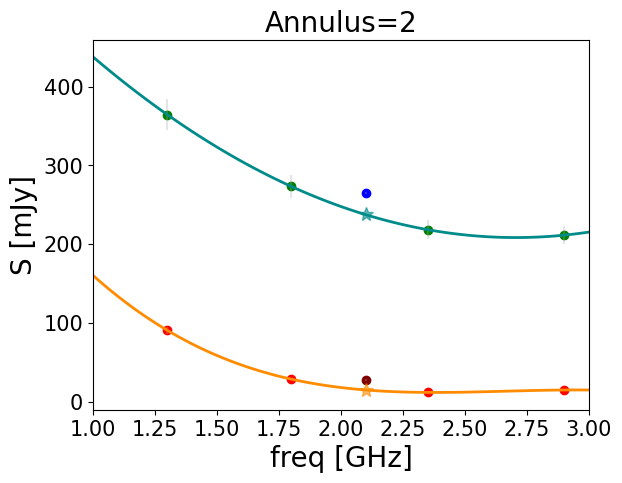}
    \caption{Examples extracted from different annuli of uncorrected (orange) vs corrected (cyan) fit to the sub-bands measurements. The stars indicate the values at 2.1\,GHz extracted from the fits. The blue dot is the corrected version of the maroon dot at 2.1\,GHz measured from the mfs maps.}
    \label{fig:spectra}
\end{figure}

\begin{figure}
    \centering
    \includegraphics[width=0.5\linewidth]{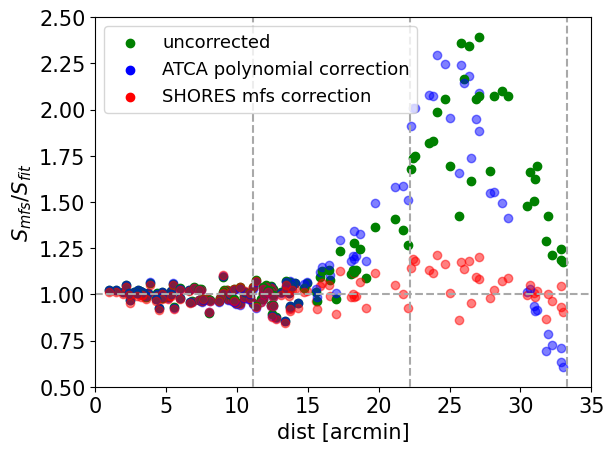}
    \caption{Ratio between the flux densities at 2.1\ GHz from the mfs map and fitted from the sub-bands as a function of the source distance from the phase center for the 546 sources with a SNR$>$5 detection in all the 4 sub-bands. The red points are the result of the application of the SHORES primary beam correction. For comparison, green dots are uncorrected, blue are the values after the polynomial ATCA primary beam correction at 2.164 GHz is applied. Vertical dashed lines correspond to the limits of the annuli. The horizontal dashed line at ratio equal to 1 is drawn for reference.
}
    \label{fig:pbcorr_calc}
\end{figure}

For construction, the SHORES constitutes a `multiple pencil beams' survey: differently from surveys that homogeneously cover a given region of the sky by combining a mosaic of pointings, our observations target non contiguous regions by sampling multiple pointings centered on specific sources, as described in previous Sections.
Imaging of each region, as mentioned, covered 3 times the FOV-FWHM per each field: to such distance, the bright sources are still detectable, increasing the surveyable area for the less dense sources. 
Therefore, the overall imaged area within three FOV from the phase center of the 27 shallow fields maps sums up to about 26 square degrees. Of course, each map is affected by the reduced sensitivity of the telescope as the distance from the phase center increases, due to the telescope primary beam response. Hence, the sensitivity is not homogeneous across the whole surveyed area, but its spatial profile is shaped by the inverse of the antenna response function, i.e. the primary beam correction. Thus, by considering a given confidence level for source detections, such profile provides an estimate for the surveyed area as a function of the flux density detection level. 

For reference, we define 3 concentric annuli in the response function of each field, corresponding to three regions of different response, from the closest to the pointing center (with best sensitivity), to the furthest (with the worst sensitivity): properties of each annulus are reported in Table \ref{tab:annuli}.

The analytic shape included in \texttt{Miriad} for the primary beam correction is truncated in distance from the phase center to the distance where it reaches $\sim10\%$ of the value in the phase center, and is considered for use within 1 FOV FWHM (see red line in  Figure \ref{fig:pbcorr}), so it does not allow for image restoration on larger areas. 
ATCA documentation\footnote{\url{https://www.narrabri.atnf.csiro.au/people/ste616/beamshapes/beamshape_16cm.html}} reports a set of polynomial fits to the primary beam response, estimated as a function of frequency in the 16\,cm band on sub-bands of 128\,MHz width, and accounting for inclinations along the legs of the secondary reflector or along the altitude-azimuth direction. The different frequencies polynomial corrections pretty much overlap within the central half FOV-FWHM (and also with the \texttt{Miriad} shape). As the distance from the phase center increases, also the difference between the beam shapes differ quite a lot. Therefore, the polynomial fit estimated on a single $128$ GHz band centered at $2.164$ GHz (the closest to our mfs nominal band center) is not enough to correct the $2.1$ GHz mfs data, as it cannot consider the variation of the beam shape as a function of frequency over the whole band, and might result in critical errors, in particular at the largest distances from the phase centers.

Therefore we directly measured the profile of the primary beam during one of our observing epochs. We pointed the ATCA antennas in 20 pointings that have a difference in the declination within the range 0-45 arcmin (moving Southerly) from the source PKSJ0537-441. This quasar is, among the brightest ($3.851\pm0.004\ $mJy) unresolved calibrators at $2.1\ $GHz, the closest to our observational setting: this target was preferred with respect to closer ones because its brightness allows a high significance detection also at the furthest distance even in the sub-bands. We measured it at different elevations across a 7hr range in Local Sidereal Time, with the same observational setup that we used in all the SHORES epochs. Calibration, imaging, source detection and flux density measurement with \texttt{BLOBCAT} followed the procedures described in previous sections, so that any effect due to the distance from the phase center in the survey sources is reasonably reproduced also for PKS0537-441, and we could compute the beam profile for each sub-band and for the mfs maps (see Figure  \ref{fig:pbcorr}). We fit it with a 8th order polynomial in the form
$$
\Pi(\nu, d)= \Sigma_{n=0}^8 a_{n}(\nu) d^n 
$$
where $d$ is the distance from the phase center in arcmin, and the parameters $a_{n}(\nu)$ for each sub-band $\nu$ are reported in Table \ref{tab:pbcorr}. The corrections, plotted in Figure  \ref{fig:pbcorr} are calculated as $1/\Pi(\nu)$ for each sub-band and the mfs maps.

We verified the difference between the estimate from the mfs map and the fit of the sub-bands spectral behavior. We corrected each sub-band measurement with its SHORES measured polynomial primary beam correction. Then we reconstructed the corrected source spectra with a third order polynomial fit and extrapolated a corrected value at 2.1 GHz, that is consistent with the overall source spectral behavior. We then compared it with the corrected point extracted from the mfs map (see some examples in Figure \ref{fig:spectra}). The mfs primary beam correction appear to be consistent with the sources spectral behavior obtained accounting for the sub-bands measurements (Figure \ref{fig:pbcorr_calc}).

In this way we can use the full surveyed area, up to 3 FOV-FWHM for each field, accounting only for the varying flux density limit as a function of distance that defines the effective area function.

\begin{figure}
    \centering
    \includegraphics[width=0.45\textwidth]{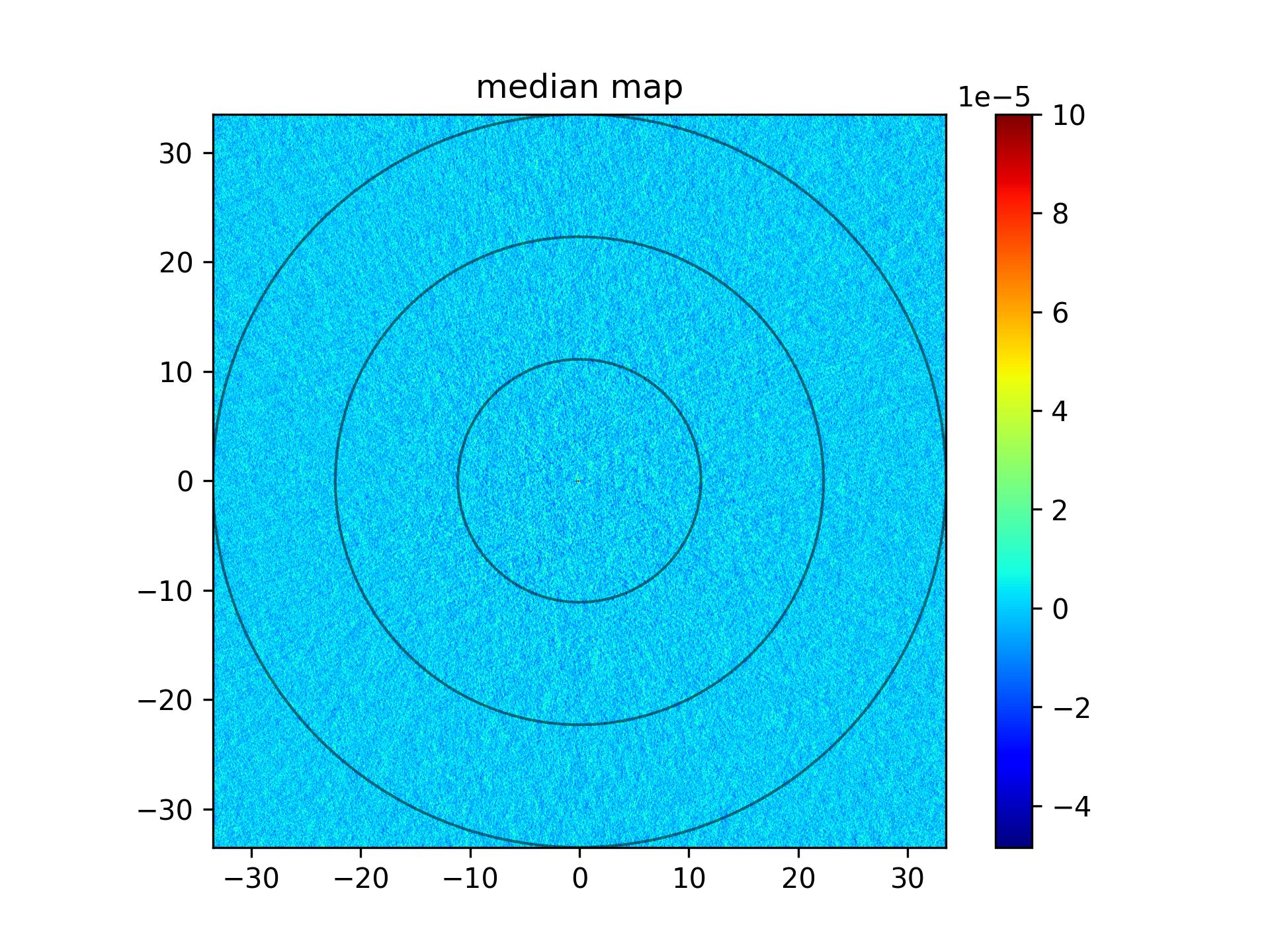}
    \includegraphics[width=0.45\textwidth]{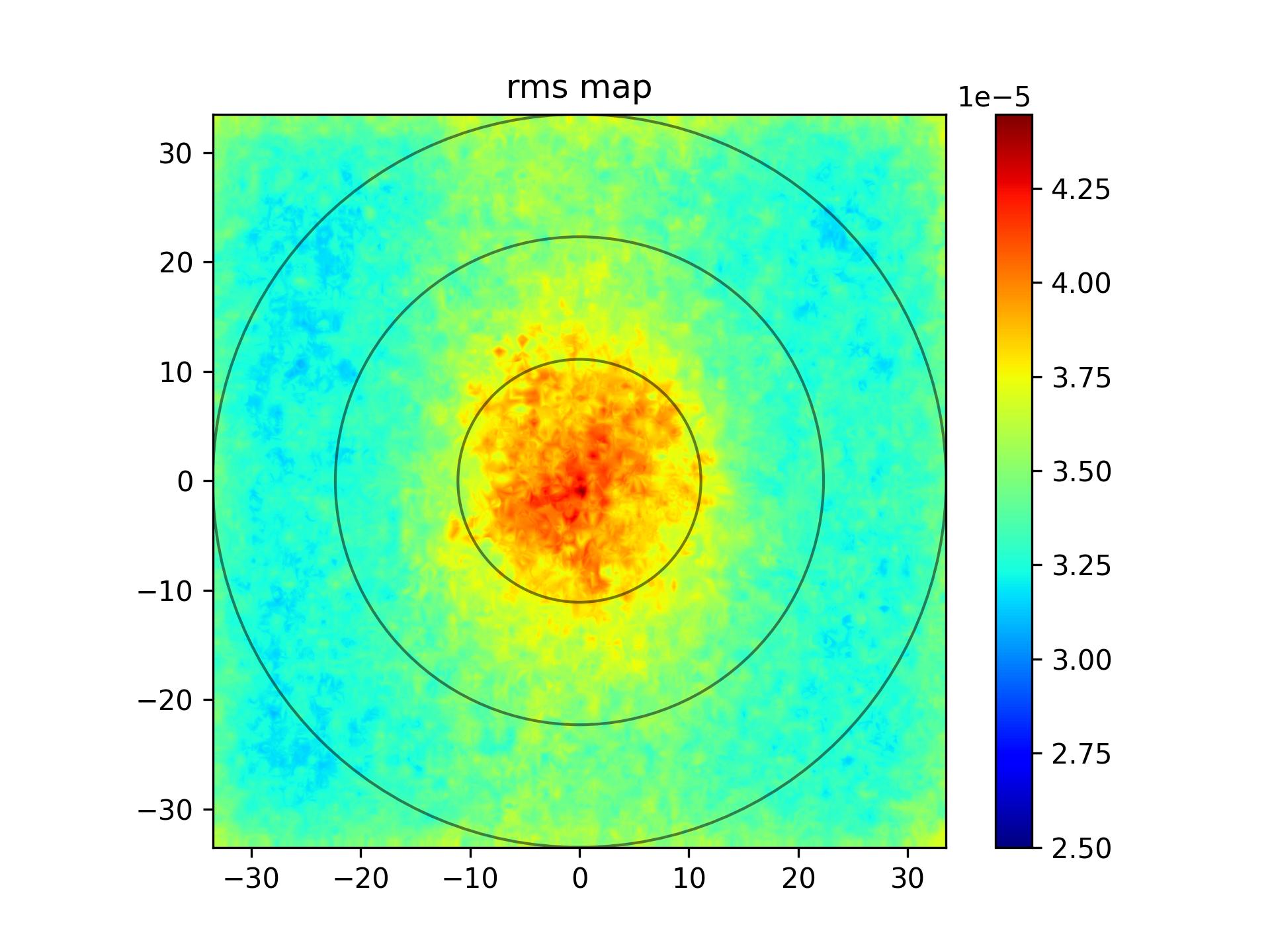}
    \caption{Map of the median of the pixels for the flux density and of the \texttt{PySE} noise maps (\textit{right panel}) of the 27 shallow fields with no primary beam correction applied.
    The annuli described in the text are overlaid in all the panels. Axis are in arcmin, colour scales in Jy.} 
    \label{fig:median_map}
\end{figure}
\begin{figure}
    \centering
    \includegraphics[width=0.50\textwidth]{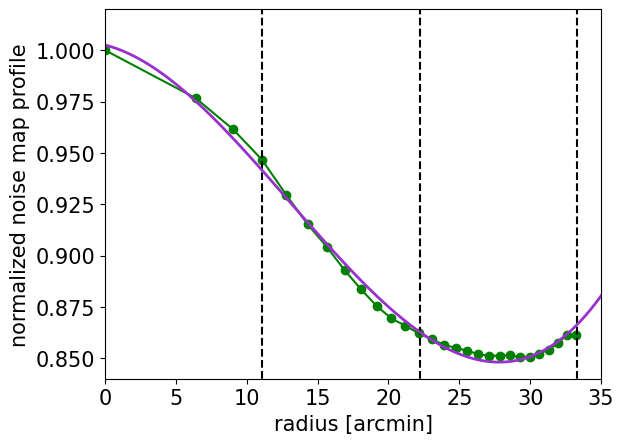}
    \caption{Normalized mean profile of the noise maps as a function of the distance from the phase center. The purple line is the fit $N(d)$ that we used to calculate the component of the noise $\sigma_{pb}(d)$. Vertical lines indicate the annuli limits.} 
    \label{fig:noise profile}
\end{figure}

\subsection{Noise profile}\label{sec:noise}

Figure \ref{fig:median_map} displays the maps of the median of the flux density and of the \texttt{PySE} noise maps of the 27 shallow fields. Because all the pointings are centered over a candidate lensed source, a median source of $223\, \mu$Jy is present at the center of the flux density median map.  

While the profile of the non-primary-beam-corrected flux density map is consistent with zero, as expected, the median profile of the noise map is not null and decreases with the distance from the phase center (see Figure \ref{fig:noise profile}). This effect seems to indicate that the value of the noise at the outer edges of our maps should be larger. The trend could be, at least partially, traced back to the effects of calibration and uncertainty in the primary beam correction as the distance from the pointing center increases. To properly and conservatively account for the associated uncertainty, we fit the noise profile $N(d)$ as a function of the distance $d$ from the phase center of each source, use it to calculate $\sigma_{pb}(d)=S_{2.1}\times [1-N(d)]$ in terms of its measured mfs flux density $S_{2.1}$, and sum it quadratically to the flux density error $\sigma_{\rm B}$ as estimated for each source from \texttt{BLOBCAT}.
Therefore, for each source the flux density error has been calculated as
$$
\sigma_{\rm{tot}}=\sqrt{\sigma_{\rm B}^2+\sigma_{pb}^2}\;.
$$

\subsection{Effective Area}\label{sec:area}

\begin{figure}
    \centering
    \includegraphics[width=0.5\textwidth]{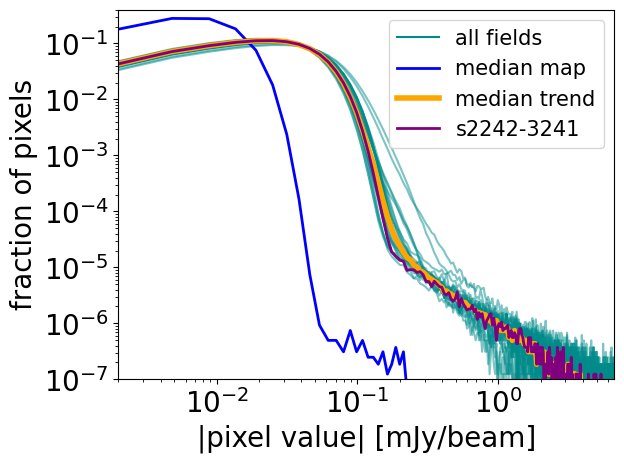}
    \caption{Fractional pixel distributions of the absolute value of total intensity maps of the 27 shallow fields and their median trend (orange line). The field s2242-3241 (purple line) is the closest to the median trend, and for this reason it is assumed as representative of our field sample. For comparison we also plot the distribution of the map obtained computing the median pixel by pixel of all the 27 maps: this is consistent with noise with the only exception of the median of the central lensed galaxies.}
    \label{fig:pixdistr}
\end{figure}
\begin{figure}
    \centering
    \includegraphics[width=0.45\textwidth]{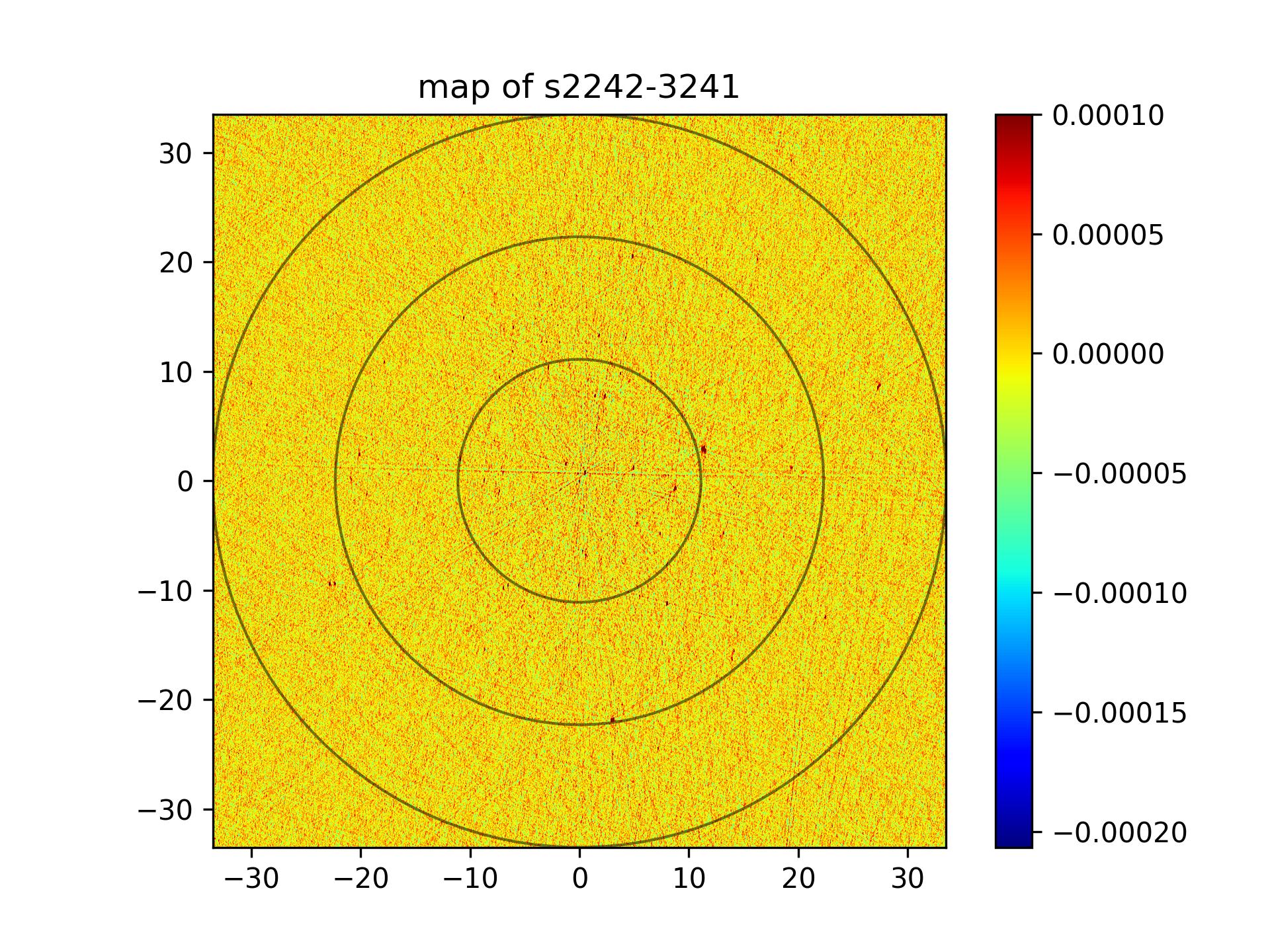}
    \includegraphics[width=0.45\textwidth]{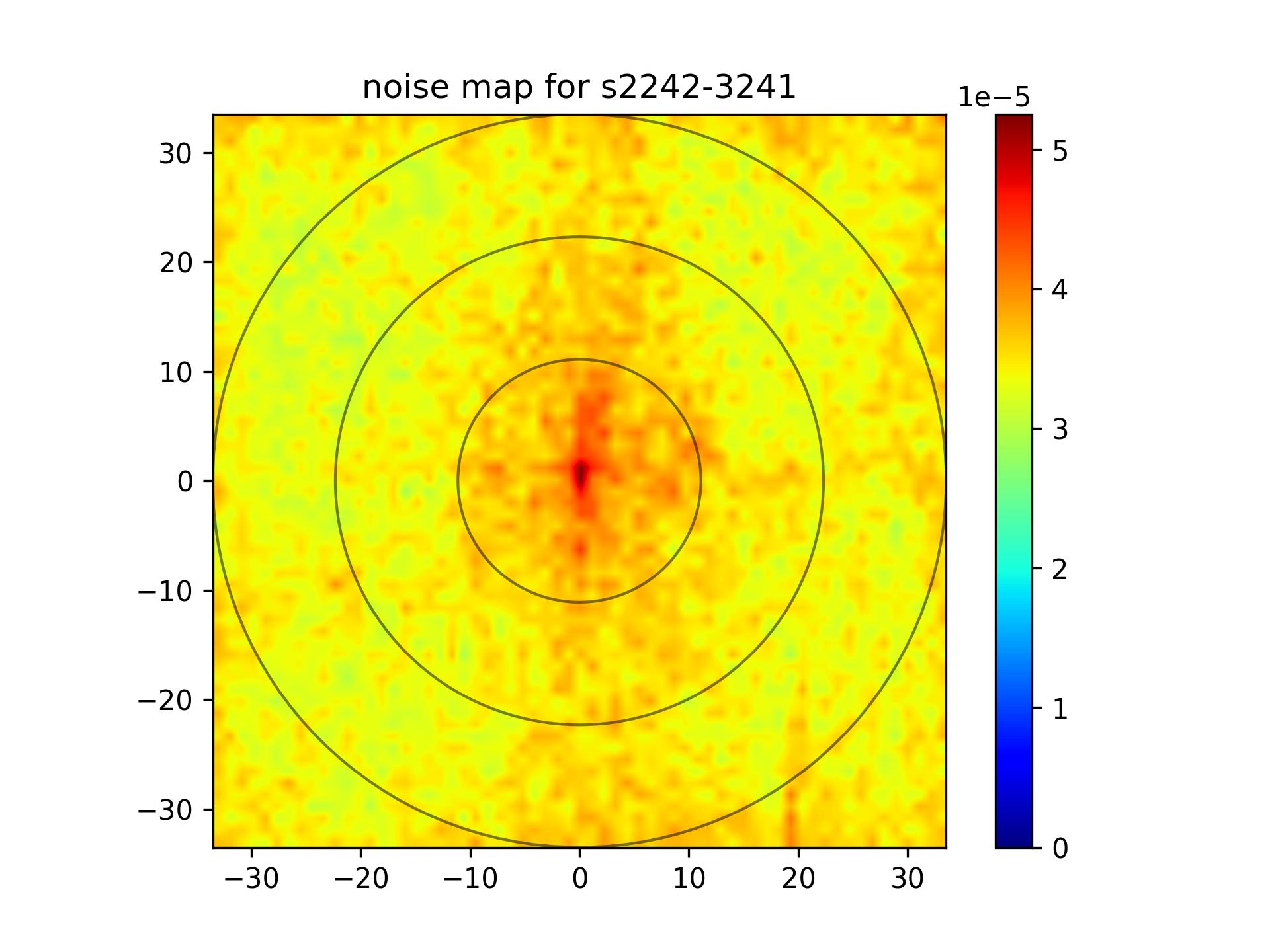}
    \caption{Map of the representative field s2242-3241 \textit{(left panel)} and its noise map as estimated by \texttt{PySE} \textit{(right)}} 
    \label{fig:s2242-3241}
\end{figure}

In each field map, the presence of sources adds a non Gaussian noise pattern component that, locally, reduces the sensitivity and might lead to false detections, in particular close to the brightest objects. On the one hand, the fields with the brightest sources present a broader pixel distribution (see Figure \ref{fig:pixdistr}), with a more evident shift towards positive flux density values, while the fields with less bright sources present a narrower distribution. The map obtained from the median of each pixel over all the fields is the narrowest since bright sources are rare in our fields, as expected. Therefore, while the median map is a representation of the median of the noise, it is not a good representation of the effects due to the presence of the sources. The field s2242-3241 is the one whose pixel distribution is closer to the median trend of all the distributions, so we chose it as nicely representative of the properties of all our fields (see Figure \ref{fig:s2242-3241}). 

\begin{figure}
    \centering
    \includegraphics[width=0.45\textwidth]{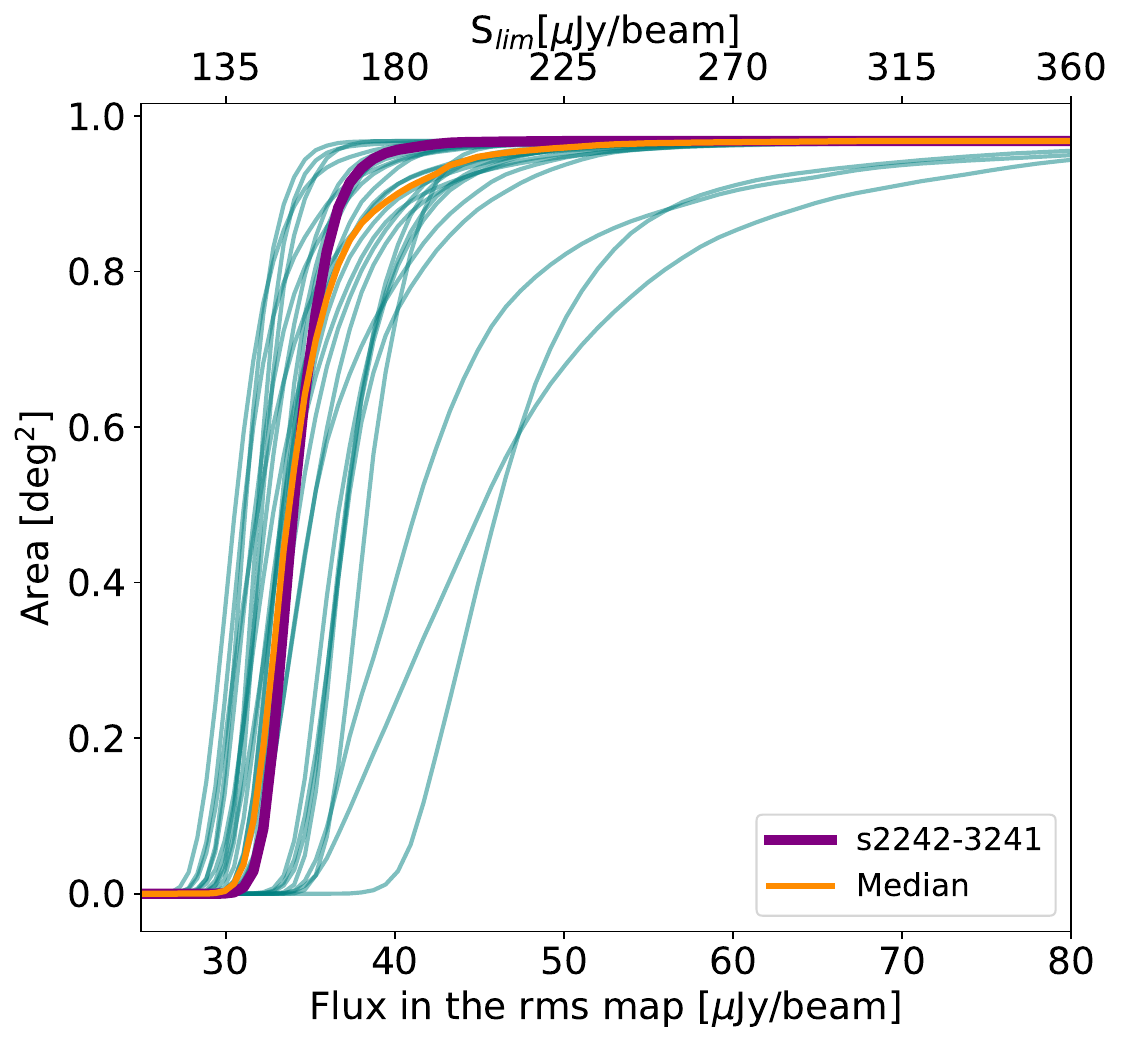}
    \includegraphics[width=0.45\textwidth]{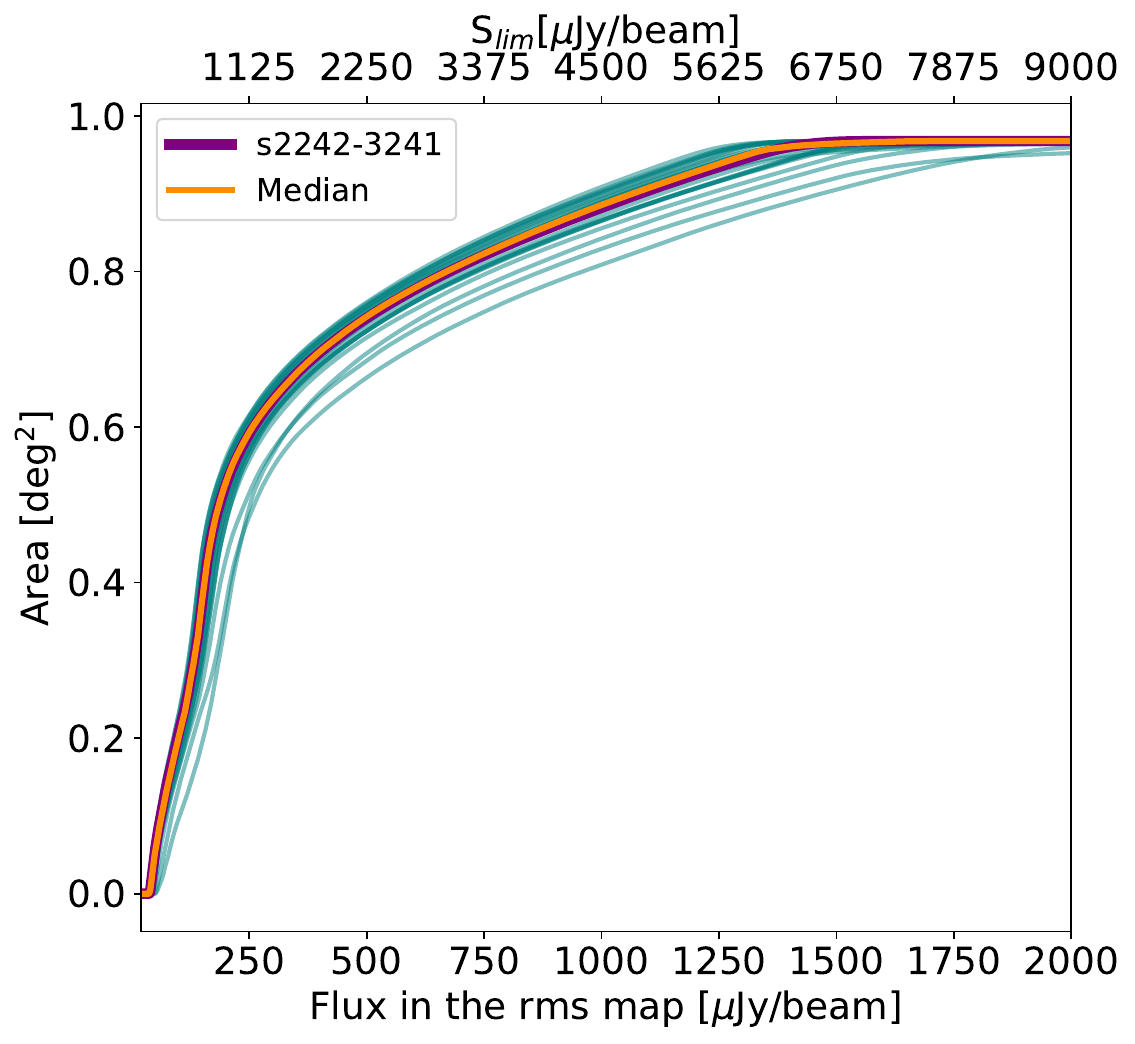}
    \caption{Cumulative area per field of the pixels with values smaller than a given value, corresponding to the effective area for each flux density limit, accounting \textit{(right panel)} or not \textit{(left panel)} for the primary beam response function: the former indicates the area on which the detections are performed, the latter indicates the area on which sources brighter than a certain noise threshold are cataloged. }
    \label{fig:eff_area}
\end{figure}

\begin{figure}
    \centering
     \includegraphics[width=0.45\textwidth]{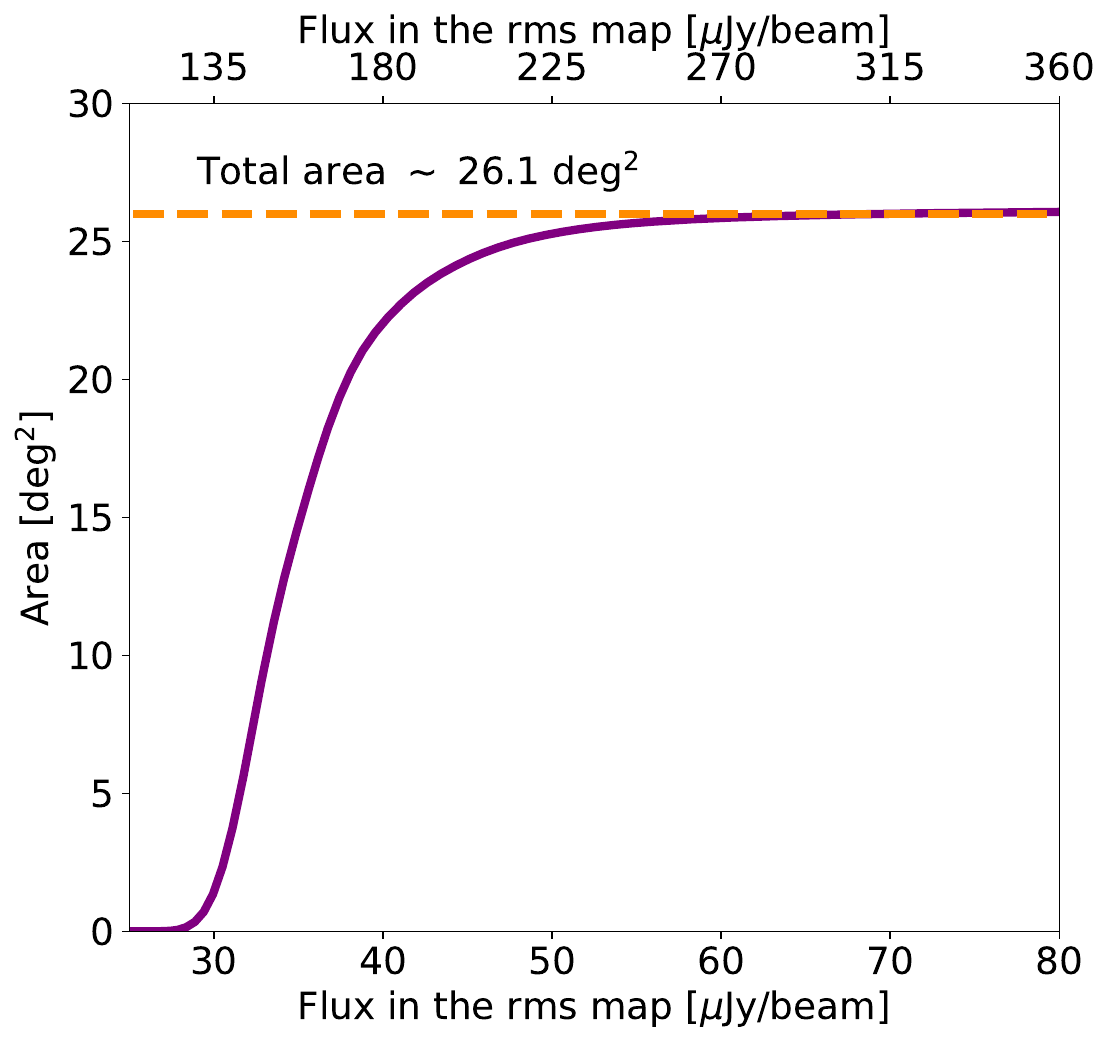} \includegraphics[width=0.45\textwidth]{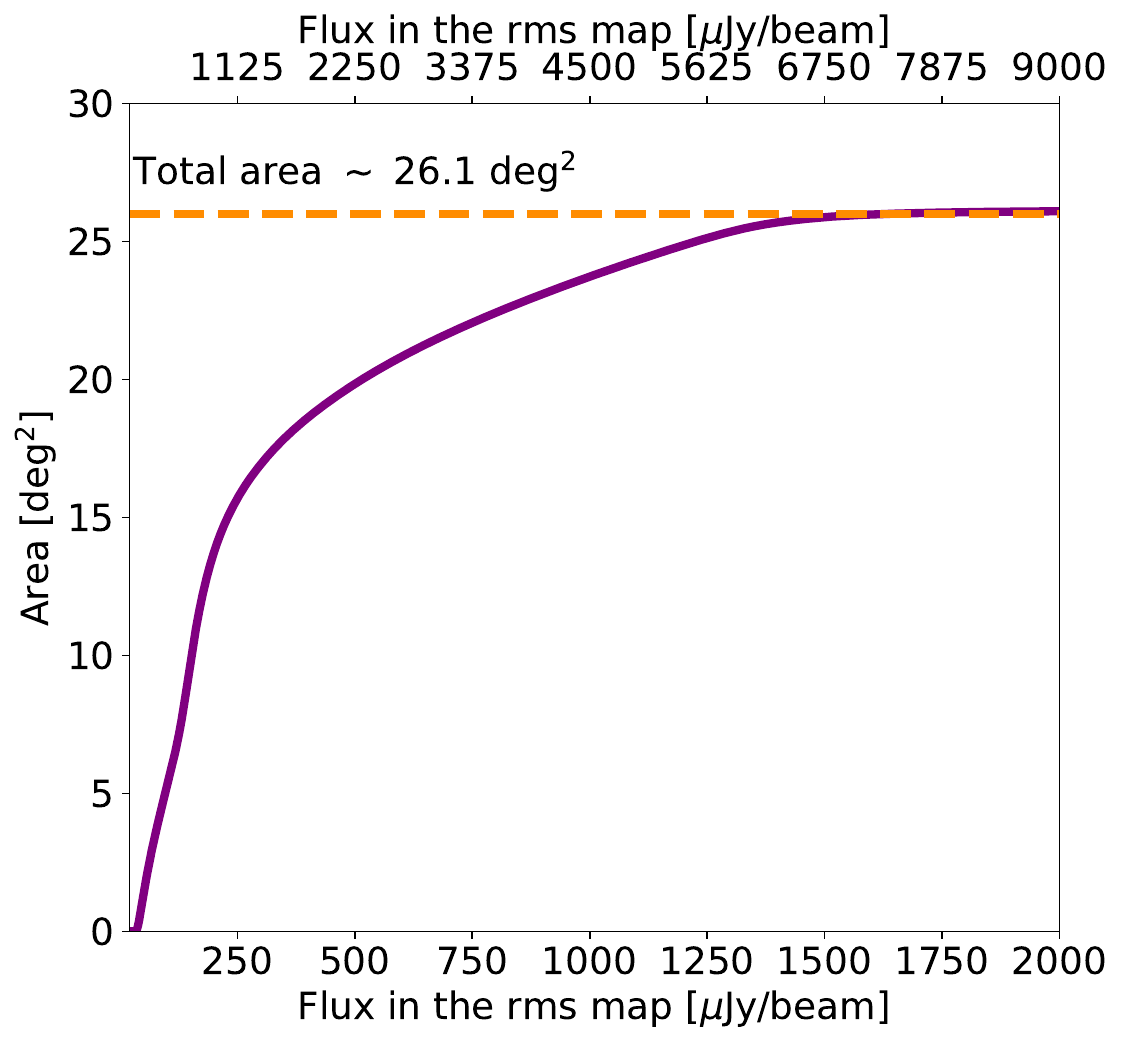}
    \caption{ Similar to Fig. \ref{fig:eff_area}, but summing all the 27 shallow survey fields.}
    \label{fig:tot_area}
\end{figure}

If we count in all the maps with no primary beam correction applied, the number of pixels with value below a certain value as a function of the value itself, we could estimate the area over which a detection can be performed per each given pixel value. The same count on the maps corrected for the primary beam response gives the effective area of the survey as a function of the sensitivity threshold, as plotted in Figure \ref{fig:eff_area}, i.e. the area above which, for each flux density level we could find sources brighter than that value. Although similar in principle, the two cumulative area distributions (in Fig. \ref{fig:tot_area}) have different behavior as a function of distance from the pointing center, since the second distribution is equivalent to the first one but convolved with the telescope response function.
This clearly shows that, even if detections with the same signal to noise ratio could be performed in all the annuli, they correspond only to brighter objects in the outer region (annulus 2), while they account also for the faintest objects in the inner regions (annulus 0). 

For reference, from this evaluation, we could collect sources above $150\ \mu$Jy with $\rm{SNR}>4.5$ in $\sim 7$ square degree (the reasons to choose the 4.5 threshold are detailed in the following Sections). 
On an effective area of the same size in the SHORES shallow fields we could identify sources brighter than 0.5\,mJy with SNR$\gtrsim 4.5$. The effective area trend with sensitivity will be considered while statistically evaluating the source counts in Section \ref{sec:SourceCounts}.

\subsection{Positional uncertainties, completeness and reliability}

\begin{figure}
    \centering
    \includegraphics[width=0.45\textwidth]{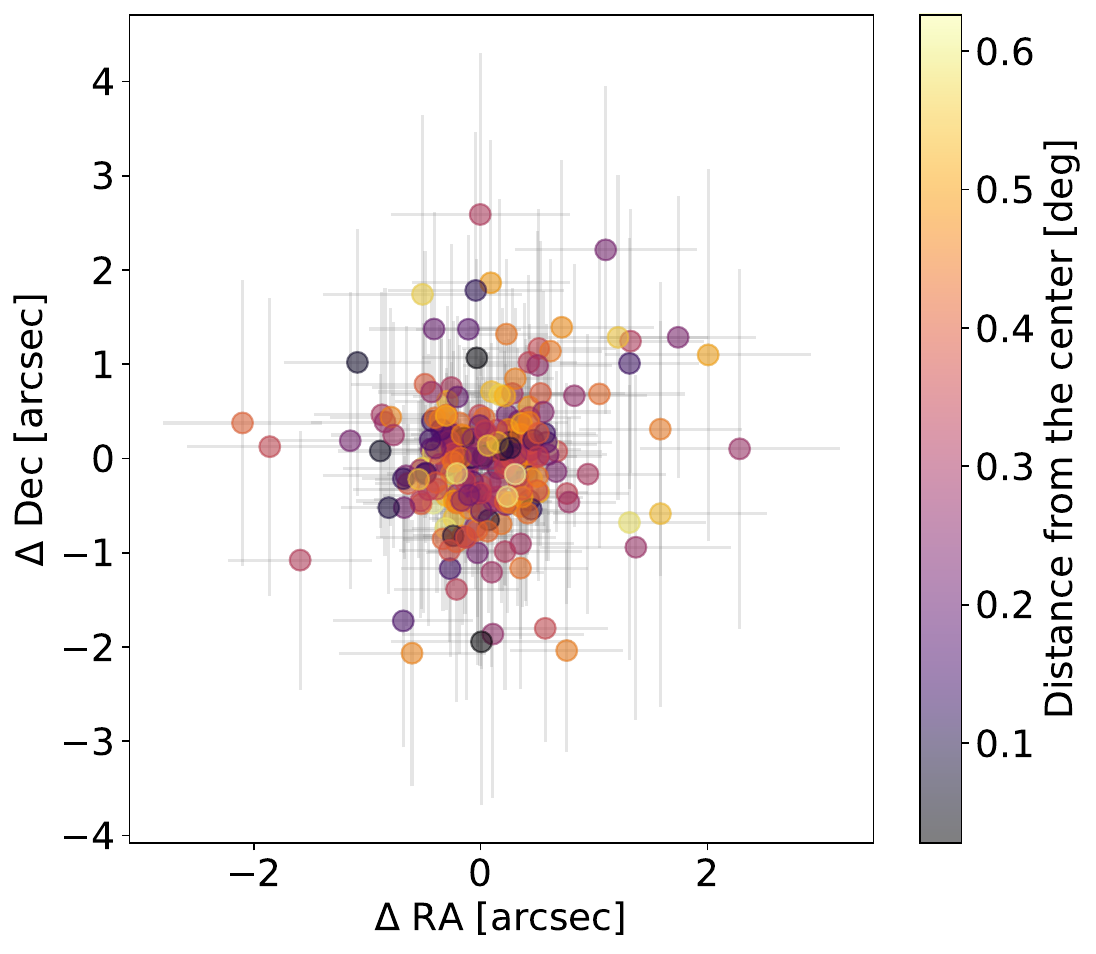}
    \includegraphics[width=0.45\textwidth]{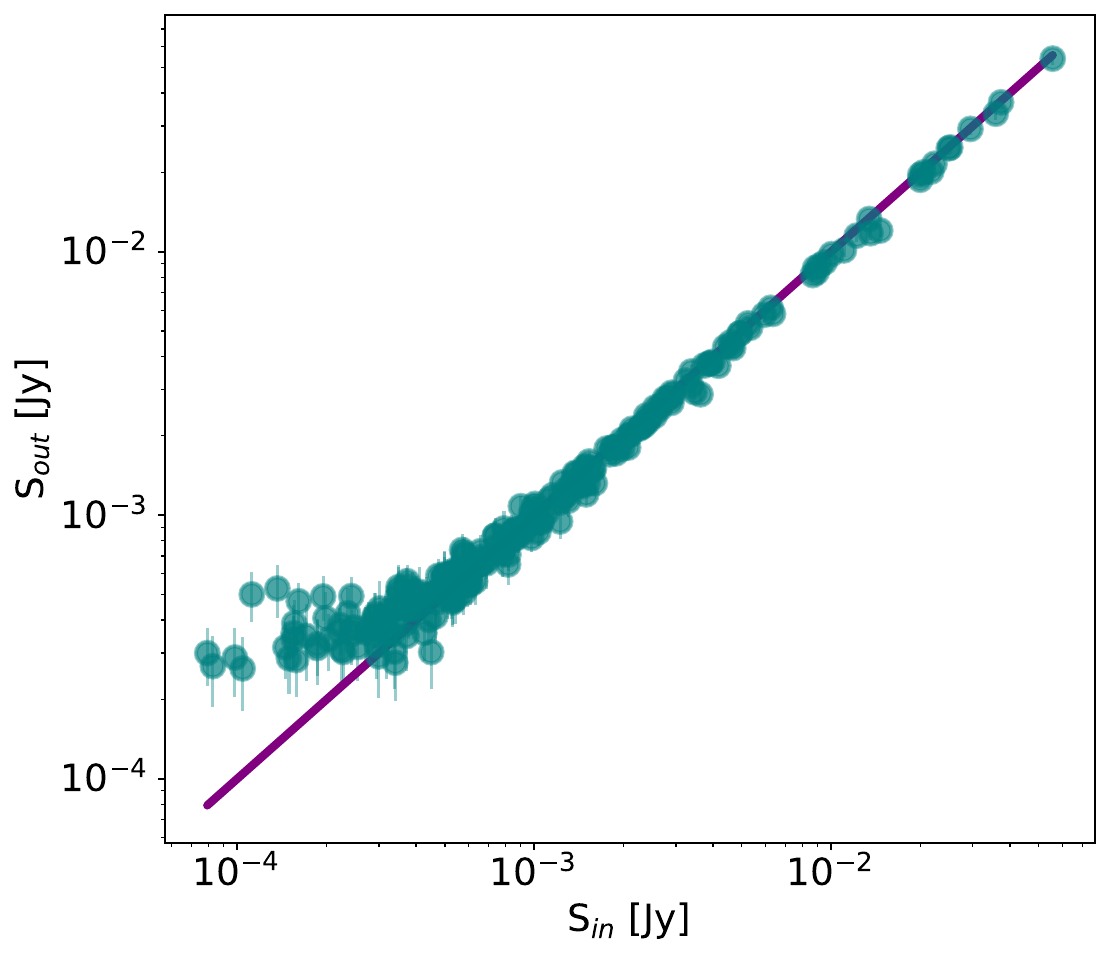}
    \caption{\textit{Left panel.} Difference in ingested and detected source position in RA and DEC for simulated sources. Colors refer to the distance from the center in deg. \textit{Right panel.} Comparison of ingested and detected flux densities for simulated sources.} 
    \label{fig:inout}
\end{figure}

To quantify the efficiency of our procedures we have tested them on simulations. As mentioned, field s2242-3241 has been identified as representative of the whole sample. We have performed 10 realizations of fake source samples, extracted accordingly to the \cite{mancuso17} source counts and randomly distributed in the field. 
To do so, we have reproduced all the real s2242-3241 field observations with the \texttt{Miriad} task \texttt{UVGEN}, generating a reliable pattern of visibilities for each realisation of the field with different simulated ingested sources. 
The same procedure to generate the real images has been applied to each simulated realization and flux densities have been extracted with \texttt{BLOBCAT}. 
We checked that the pixel distribution of a realization with point sources flux densities extracted from the source counts of 2.1 GHz population and the distribution of the original field s2242-3241 cannot be statistically distinguished, according to a Kolmogorov-Smirnov test.

We verified the position uncertainties  of our detections by comparing the ingested and the detected positions of the sources (see Figure \ref{fig:inout}). The median positional error is $(\delta {\rm RA}, \delta {\rm dec}) \approx (0.036'', 0.032'')$, a tiny fraction of the pixel size ($1''$). Uncertainty, of course, increases as the source flux density becomes fainter. No dependence on the distance from the pointing center is clearly visible. 
By comparing the input and output flux density (see Figure \ref{fig:inout}) for a sample of 1000 point sources in 10 realizations we confirmed that there are no significant differences in the different annuli. A hint of excess in the output flux density with respect to the input flux density could be observed for sources below $0.5$ mJy, if only sources detected to a $3$ SNR level are considered: this is likely due to Eddington bias.   

\begin{figure}
    \centering
    \includegraphics[width=0.45\textwidth]{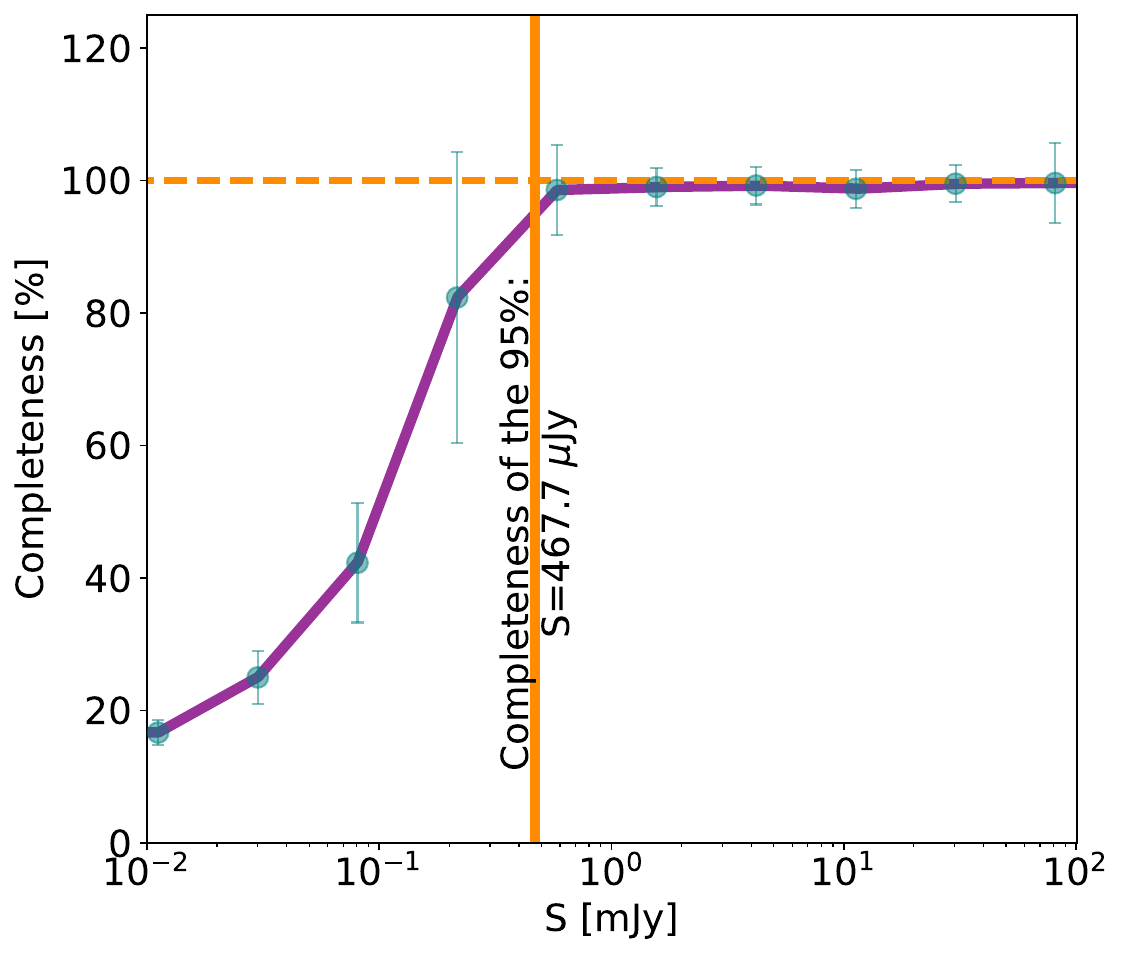}
    \includegraphics[width=0.45\textwidth]{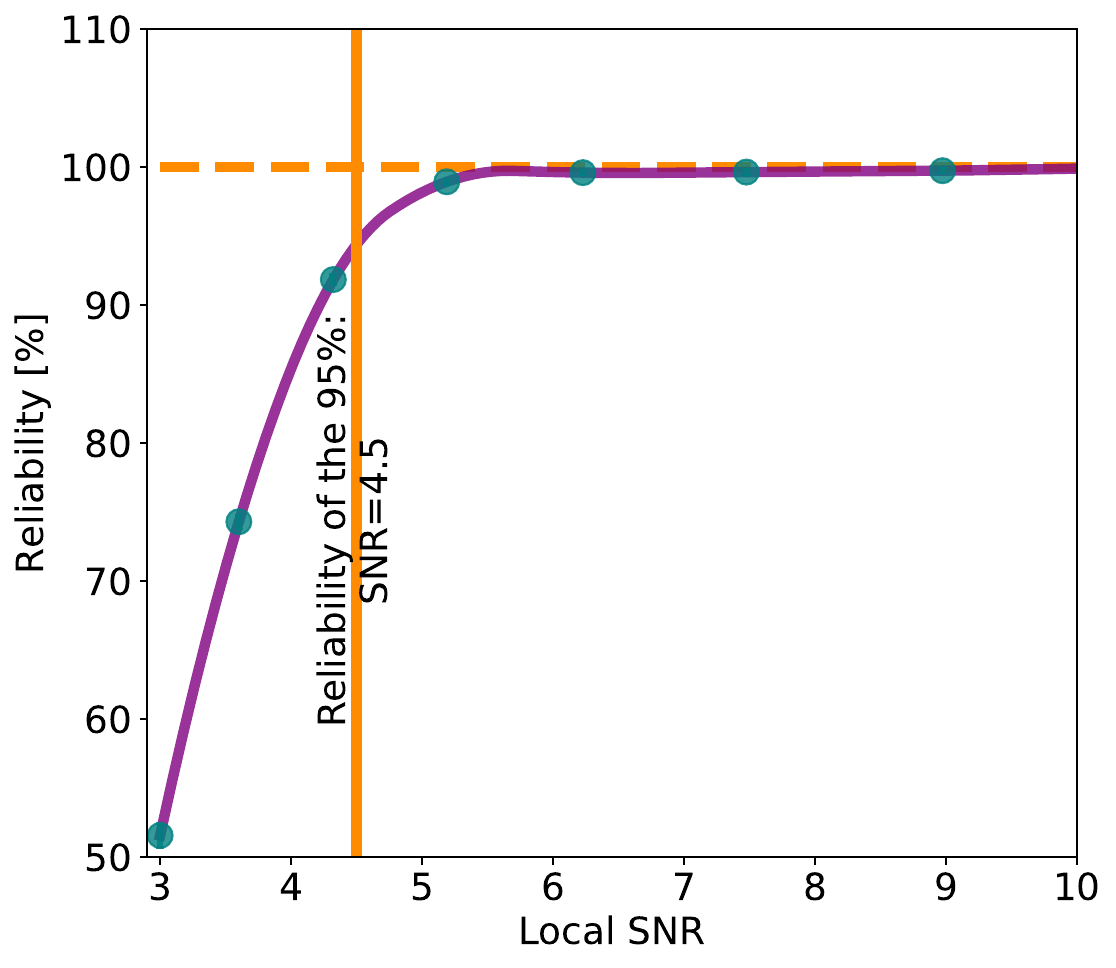}
    \caption{Completeness (\textit{left panel}) and reliability (\textit{right panel}) of our simulations. } 
    \label{fig:completeness}
\end{figure}

To evaluate the effectiveness of our survey, we determined the fraction of detected sources as a function of flux density, which defines the survey's completeness. This completeness needs to be considered when calculating radio source counts. To perform the simulation, we injected 10000 point sources of flux density extracted according to the radio number counts by \cite{mancuso17}, one at a time, into random positions on the inverse map of s2242-3241 using the \texttt{Miriad} task \texttt{IMGEN}. \texttt{BLOBCAT} was then executed on those positions with the same input parameters that were used to detect the real radio sources in the survey. Thus, we estimated the ratio of the injected sources that \texttt{BLOBCAT} was able to detect as ($N_{\rm det}/N_{\rm in}$). The completeness percentage is defined as $100\times N_{\rm det}/N_{\rm in}$. Its Poissonian uncertainty is computed as $100\times \sqrt{N_{\rm det}}/N_{\rm in}$.

Then, we calculated the percentage of false detections at a given SNR to establish the threshold at which a detected object can be considered a true source rather than an artifact. For this purpose, we executed \texttt{BLOBCAT} blindly on the inverted maps of the survey. Consequently, we derived the ratio between detections on the negative maps and those on the actual maps for various SNRs ($N_{\rm neg}/N_{\rm pos}$). This ratio represents the false detection rate ${\rm FDR} \approx N_{\rm neg}/N_{\rm pos}$, and $R = (1-{\rm FDR})\times 100$\% defines the reliability of the survey. Thus, we found that a detection threshold of SNR$\gtrsim 4.5$ is required to ensure that the probability of a detection to be an artifact is less than $5\%$ (see Figure \ref{fig:completeness}).

\begin{figure}
    \centering
    \includegraphics[width=0.5\textwidth]{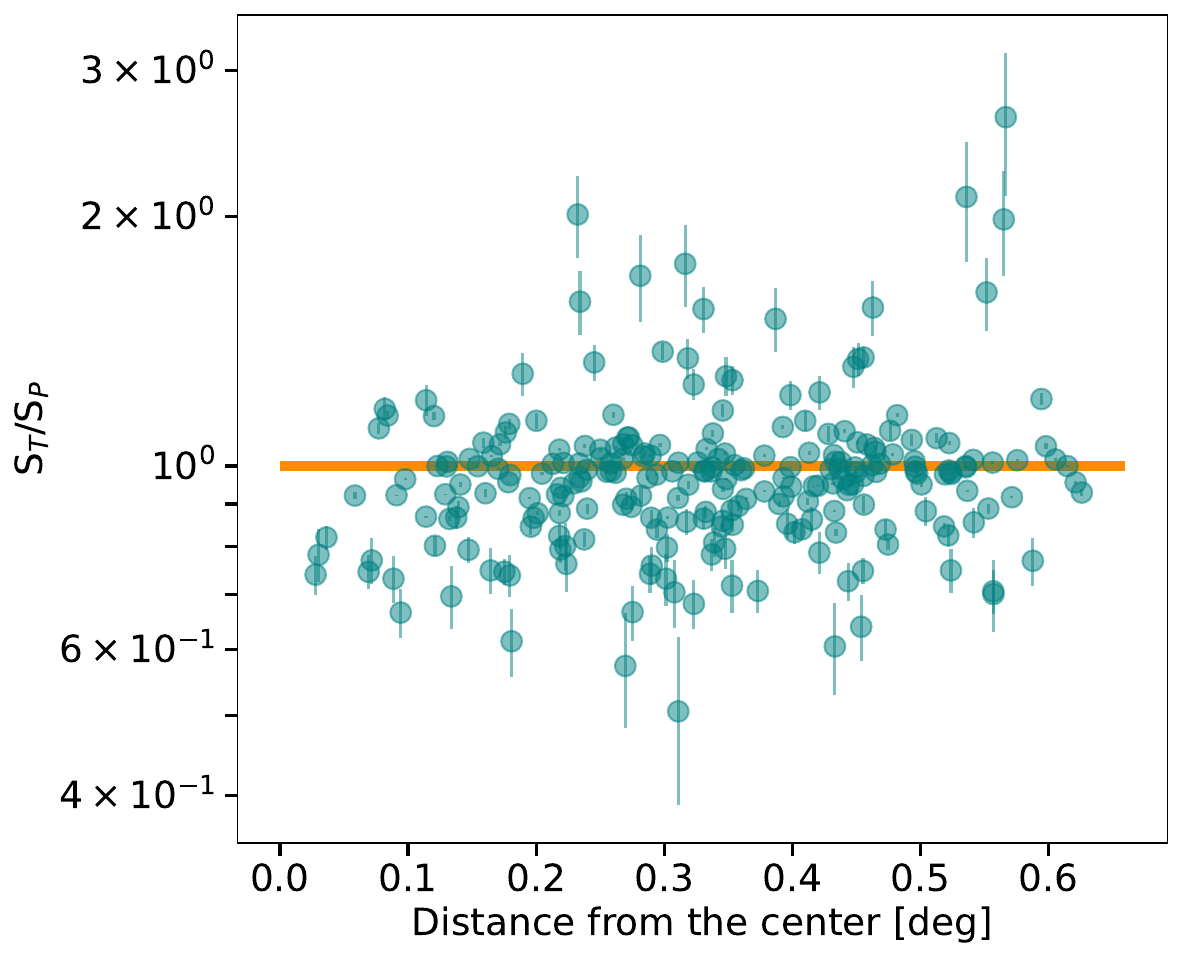}
    \caption{Ratio of the integral vs peak flux densities for simulated point sources at varying distance from the phase center.} 
    \label{fig:fpeakfint}
\end{figure}

\begin{figure}
    \centering
    \includegraphics[width=0.5\textwidth]{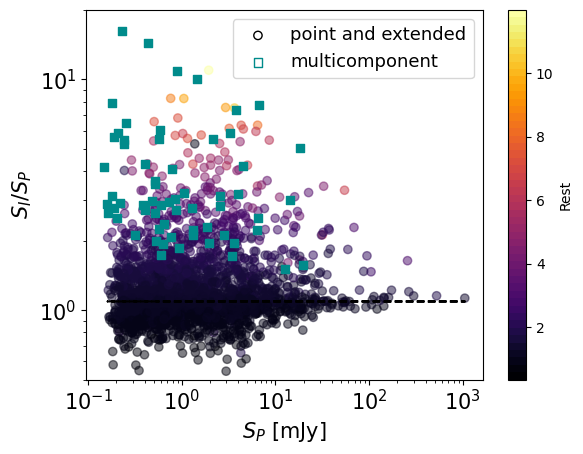}
    \caption{Ratio between total flux and peak flux as a function of the peak flux for the SHORES sources. Extended and point sources are colored according to the \texttt{BLOBCAT} $R_{\rm est}$ parameter value (the size of the candidate source in units of numbers of Gaussian elements with peak equal to the peak flux density). Multi-component sources identified in our visual inspection are highlighted in cyan squares. The median for the point sources (dashed black line) is consistent with 1 as also demonstrated by our simulations.}
    \label{fig:st_sp}
\end{figure}

Then, by comparing the ratio of the recovered integrated and peak flux densities with the signal to noise ratio of each simulated point source (Figure \ref{fig:fpeakfint}) we can estimate the effect due to bandwidth smearing \citep[i.e. the underestimation of the peak flux due to the varying synthesized beam size as a function of frequency across the observed bandwidth,][]{condon98, bondi08, bridle99}. We find that the ratio of the total and peak flux is consistent with no significant variation in the different annuli, as expected as \texttt{BLOBCAT} performs bandwidth smearing correction. The analysis of the $R_{\rm est}$ parameter, the size of the candidate source in units of numbers of Gaussian elements with peak equal to the peak flux density, indicates that $R_{\rm est} = 2$ could be reached for simulated point sources: we will use this threshold to distinguish genuinely extended and point sources \citep[similarly to what was done by][see also Figure \ref{fig:st_sp} for the catalogued sources]{butler18}. 

\section{The SHORES 2.1 GHz catalog}\label{sec:catalog}

The SHORES 2.1 GHz catalog of the $27$ shallow field is defined by the detections with \texttt{BLOBCAT}.

\texttt{BLOBCAT} detections with size $R_{\rm est}>2$ are considered resolved and for them the flux density is assumed to be equal to the \texttt{BLOBCAT} corrected integral flux (e.g. $S\_int\_CB$). They will be tagged with an 'e' in the catalog. 
Similarly, for point sources the flux density is assumed to be equal to the \texttt{BLOBCAT} corrected peak flux (e.g. $S\_p\_CBBWS$), corrected with the SHORES primary beam correction. In all the cases errors are defined by the detected flux error.
As anticipated, a total of 13688 blobs have been detected with $\rm{SNR}\gtrsim 3$ and 2649 with $\rm{SNR}\gtrsim 4.5$.

A visual inspection has been performed over all the detections made with \texttt{BLOBCAT} above $3\,\sigma_B$ to identify and remove detections that are lying on positions clearly contaminated by emission from bright sources, for which, even if the detection might be a real source, the determined flux is surely contaminated and thereby wrong. This procedure preserved 4067 blobs detected above $\rm{SNR}\gtrsim 4$. 

A second visual inspection of the $\rm{SNR}\gtrsim 4$ blobs has been performed to match detections that have a strong probability to be multiple components of the same source (e.g. lobes of the same AGN): 69 similar cases have been classified, and flagged with a `m' in the catalog. For them a region enclosing the whole source has been determined on the map and the flux has been defined as the integral of the pixel values enclosed in the region. As per error on the flux estimation we have been using the maximum integral flux error of the components. 

Finally we preserved a sample of 2294 sources with $\rm{SNR}\gtrsim 4.5$, 95\% complete above $0.5$ mJy and $95\%$ reliable. Of them, 376 are extended and 62 have multiple components. A summary of the properties of the SHORES shallow fields and the number of $\rm{SNR}\gtrsim 4.5$ detections in each of them are collected in Table \ref{tab:fields}.

\begin{sidewaystable}
\centering
  \resizebox{\columnwidth}{!}{%
  \begin{tabular}{l|l|c|c|c|c|c|c|c|c|c|c|c}
    \hline
    Field & Central Coordinates & Synthesized Beam & $\sigma_{B}$ & A$_{\text{eff}}^{\text{det}}$ & A$_{\text{eff0}}^{\text{pb}}$  & S$_{\text{median0}}^{\text{pb}}$ & A$_{\text{eff1}}^{\text{pb}}$  & S$_{\text{median1}}^{\text{pb}}$ & A$_{\text{eff2}}^{\text{pb}}$  & S$_{\text{median2}}^{\text{pb}}$ & \# Sources & \# Sources H-ATLAS \\
    & RA dec & [arcsec$^2$] & mJy& [deg$^2$] & [deg$^2$] & mJy & [deg$^2$] & mJy & [deg$^2$] & mJy  & at 4.5$\sigma$ & at 4.5$\sigma$  (match radius 25 arcsec) \\
    \hline
    s0000-3340 & 00h00m07.4s -33d41m59.0s & 27.4 & 38.0 & 1.0 & 0.08& 0.27 & 0.27& 1.6 & 0.24 & 3.7 & 105 & 21\\
    s0007-3520 & 00h07m22.1s -35d20m14.0s & 26.6 & 37.0 & 1.1 & 0.08 & 0.26 & 0.26 & 1.47& 0.20& 3.46& 109 & 23 \\
    s0026-3417 & 00h26m24.8s -34d17m37.0s & 29.7 & 34.0 & 1.0 & 0.1& 0.34 & 0.27 & 1.42 & 0.46 & 3.76 & 95 & 17 \\
    s0032-3037 & 00h32m07.7s -30d37m24.0s & 32.5 & 33.0 & 1.1 &  0.07 & 0.24 & 0.28 & 1.5 & 0.54 & 6.04 & 69 & 17\\
    s0048-3031 & 00h48m53.2s -30d31m09.0s & 32.6 & 35.0 & 1.1 & 0.08 & 0.25 & 0.29 & 1.72 & 0.51 & 1.72 & 84 & 23 \\
    s0102-3117 & 01h02m50.8s -31d17m23.0s & 32.3 & 33.0 & 1.0 & 0.10 & 0.29& 0.26 & 1.27& 0.18 & 3.10 & 105 & 26 \\
    s0120-2824 & 01h20m46.4s -28d24m03.0s & 33.2 & 35.0 & 1.0 & 0.10 & 0.29 & 0.30 & 1.84 & 0.43 & 3.83 & 63 & 21\\
    s0124-2814 & 01h24m07.3s -28d14m34.0s & 32.6 & 37.0 & 1.0 & 0.10 & 0.32 & 0.26 & 1.4 & 0.27 & 3.64 & 79 & 16 \\
    s0124-3105 & 01h24m15.9s -31d05m00.0s & 33.7 & 32.0 & 1.1 & 0.11 & 0.31 & 0.26 & 1.25 & 0.16 & 2.86 &97 &23 \\
    s0130-3055 & 01h30m04.0s -30d55m13.0s & 34.0 & 31.0 & 1.1 & 0.09 & 0.24 & 0.26 & 1.18 & 0.27 & 3.00 &76 & 17 \\
    s0132-3309 & 01h32m39.9s -33d09m06.0s & 31.9 & 31.0 & 1.1 & 0.11 & 0.34 & 0.25 & 1.11 & 0.54 & 4.32 & 82 &19\\
    s0138-2818 & 01h38m40.4s -28d18m55.0s & 33.4 & 34.0 & 1.1 & 0.11 & 0.32 & 0.26 & 1.27 & 0.22 & 3.16 & 84 & 22 \\
    s0139-3214 & 01h39m51.9s -32d14m46.0s & 33.0 & 33.0 & 1.1 & 0.10 & 0.32 & 0.27 & 1.36 & 0.54 & 6.61 & 80 & 17\\
    s2237-3058 & 22h37m53.8s -30d58m28.0s & 32.1 & 34.0 & 1.1 & 0.09 & 0.29 & 0.25 & 1.23 & 0.31 & 3.38 & 74 & 18 \\
    s2242-3241 & 22h42m07.2s -32d41m59.0s & 33.2 & 34.0 & 1.0 & 0.08 & 0.24 & 0.25 & 1.25 & 0.16 & 3.13 & 75 & 13\\
    s2248-3358 & 22h48m05.3s -33d58m20.0s & 32.5 & 32.0 & 1.1 & 0.07 & 0.22 & 0.29 & 1.49 & 0.45 & 3.5 & 92 & 19\\
    s2252-3136 & 22h52m50.7s -31d36m57.0s & 34.5 & 32.0 & 1.1 & 0.09 & 0.28 & 0.28 & 1.40 & 0.25 & 3.05 & 85 & 23 \\
    s2258-2951 & 22h58m44.7s -29d51m24.0s & 36.4 & 32.0 & 1.0 & 0.08 & 0.23 & 0.28 & 1.39 & 0.33 & 3.34 & 91 & 15\\
    s2305-3310 & 23h05m46.2s -33d10m38.0s & 34.3 & 31.0 & 1.0 & 0.07 & 0.22 & 0.25 & 1.12 & 0.17 & 2.88 & 99 & 34 \\
    s2308-3438 & 23h08m15.5s -34d38m01.0s & 32.7 & 34.0 & 1.1 & 0.11 & 0.38 & 0.26 & 1.27 & 0.53 & 4.44 & 95 & 19\\
    s2324-3239 & 23h24m19.8s -32d39m26.0s & 24.8 & 37.0 & 1.1 & 0.11 & 0.42 & 0.25 & 1.34 & 0.17 & 3.39 & 62 & 15\\
    s2325-3022 & 23h25m31.3s -30d22m35.0s & 26.7 & 37.0 & 1.1 & 0.10 & 0.360 & 0.27 & 1.59 & 0.20 & 3.45 & 85 & 23\\
    s2326-3426 & 23h26m23.0s -34d26m42.0s & 29.3 & 34.0 & 1.0 & 0.08 & 0.25 & 0.25 & 1.21 & 0.37 & 3.56 & 99 &19  \\
    s2329-3217 & 23h29m00.6s -32d17m44.0s & 32.2 & 32.0 & 1.0 & 0.08 & 0.23 & 0.25 & 1.14 & 0.12 & 2.87 & 81 & 23 \\
    s2343-3517 & 23h43m57.7s -35d17m23.0s & 25.8 & 41.0 & 1.1 & 0.10 & 0.34 & 0.31 & 2.24 & 0.54 & 17.61 & 76 & 20\\
    s2344-3039 & 23h44m18.1s -30d39m36.0s & 32.7 & 43.0 & 1.0 & 0.10 & 0.66 & 0.31 & 2.53 & 0.54 & 2.53 & 72 & 18\\
    s2358-3232 & 23h58m27.6s -32d32m44.0s & 27.7 & 46.0 & 1.1 & 0.11 & 0.46& 0.32& 2.7 & 0.54 & 10.68 & 83 &  19\\
    \hline
  \end{tabular}%
  }
    \caption{Properties of the survey shallow fields at 2.1 GHz.  A$_{\text{eff}}$ columns indicate the effective area of the detections map, i.e. not primary beam corrected, A$_{\text{eff}}^{det}$, and for each annulus (0, 1, 2) of the primary beam corrected maps, A$_{\text{eff i}}^{\text{pb}}$. In the same way, S$_{\text{median i}}^{\text{pb}}$ indicates the flux corresponding to 4.5 $\times$ RMS$_{median}$ over each annulus. \# Source and \# Sources H-ATLAS indicate all the sources detected at SNR$>$4.5$\sigma$ and having a counterpart in the H-ATLAS catalog respectively. }\label{tab:fields}
\end{sidewaystable}

\begin{sidewaystable}
\centering
  \resizebox{\columnwidth}{!}{%
  \begin{tabular}{l|l|c|c|c|c|c|c|c|c|c|c|c|c}
    \hline
SHORES\_id               & field      &   ra[deg] & dec[deg] &annulus &Spbcorr &flux[mJy] &fluxerr[mJy] &flags &ramin &ramax &decmin &decmax &H-ATLAS\_id\\
\hline
SHORESJ000001.1-333617.0 & s0000-3340 &  0.00449 &-33.60472 &0 & 1.10  &  0.804 &   0.063 &pAP &.         &.         &.         &.         &. \\
SHORESJ000002.8-332246.0 & s0000-3340 &  0.01187 &-33.37944 &1 & 4.36  &  2.533 &   0.217 &pAP &.         &.         &.         &.         &. \\
SHORESJ000007.1-334103.0 & s0000-3340 &  0.02950 &-33.68417 &0 & 1.00  &  2.047 &   0.119 &pAP &.         &.         &.         &.         &HATLASJ000007.5-334100\\ 
SHORESJ000007.2-331613.0 & s0000-3340 &  0.02984 &-33.27027 &2 &10.91  &  2.253 &   0.481 &p.P &.         &.         &.         &.         &. \\
SHORESJ000008.0-334020.0 & s0000-3340 &  0.03358 &-33.67212 &0 & 1.00  &  1.274 &   0.077 &eAP &  0.03183 &  0.03517 &-33.67417 &-33.67000 &HATLASJ000008.2-334018\\ 
SHORESJ000008.9-334135.0 & s0000-3340 &  0.03771 &-33.69283 &0 & 1.00  &  0.463 &   0.050 &eAP &  0.03651 &  0.03918 &-33.69417 &-33.69167 &HATLASJ000009.1-334134 \\
SHORESJ000009.2-335902.0 & s0000-3340 &  0.03838 &-33.98375 &1 & 4.24  &  2.009 &   0.196 &eAP &  0.03720 &  0.03954 &-33.98528 &-33.98222 &. \\
SHORESJ000011.3-333653.0 & s0000-3340 &  0.04718 &-33.61472 &0 & 1.07  &  0.262 &   0.048 &pAP &.         &.         &.         &.         &HATLASJ000011.4-333654 \\
SHORESJ000014.6-323607.2 & s2358-3232 &  0.06054 &-32.60144 &2 & 8.32  &  4.257 &   0.445 &eA. &  0.05947 &  0.06147 &-32.60338 &-32.59950 &. \\
SHORESJ000015.7-331220.0 & s0000-3340 &  0.06536 &-33.20555 &2 &17.92  &  4.173 &   0.807 &pAP &.         &.         &.         &.         &. \\
SHORESJ000015.7-341016.5 & s0000-3340 &  0.06563 &-34.17126 &2 &19.37  & 65.644 &   1.941 &mAP &  0.05416 &  0.07792 &-34.18111 &-34.15771 &. \\
\hline
  \end{tabular}%
}
\caption{Extract of the SHORES sources catalog.}\label{tab:src_catalog}
\end{sidewaystable}

The first lines of the catalog of the SHORES sources are reported in Table \ref{tab:src_catalog}. The full catalogue is available as additional material for the present paper and on the SHORES website\footnote{\url{https://sites.google.com/inaf.it/shores}}. 

The listed fluxes have been corrected with the SHORES primary beam correction as described in Section \ref{sec:pbcorr}.
Their noise has been corrected to account for the effects of the distance from the phase center as described in Section \ref{sec:noise}. Columns include: 
\begin{description}
    \item[Source name] in the IAU format SHORESJhhmmss.s-ddmmss.s built up on the basis of the coordinate of the peak flux density pixel for point sources or on the basis of the centroid coordinate for extended or multiple component sources;
    \item[ID of the field] to which the source belongs according to Table \ref{tab:fields};
    \item[RA and dec of J2000 position] of the source (the same used for the name) in degrees;
    \item[annulus] of the field maps to which the source belongs, defined as "0" within a 0.5 FWHM radius from the field center, "1" within the distance range 0.5 and 1 FWHM from the field center, "2" between 1 and 1.5 FWHM from the field center;
    \item[SHORES primary beam correction]: multiplicative term estimated as described in Section \ref{sec:pbcorr};
    \item[primary beam corrected fluxes and their error]: in case of point sources it reports the peak flux density extracted with \texttt{BLOBCAT} (with all the corrections applied); in case of extended sources it is the integrated flux over the source region extracted with \texttt{BLOBCAT}; for multi-component sources it is the integrated flux extracted from visual inspection of the maps by defining the source footprint with the \texttt{DS9} tools. The errors account for the flux density uncertainty, calibration uncertainty and effects of distance from the phase center;
    \item[flags]: 3 digits defined as: 1) either "p" for point-like sources, "e" for extended sources, as determined by \texttt{BLOBCAT}, or "m" for multi-component sources identified during visual inspection; 2) "A" if the source is detected by \texttt{AEGEAN} or "." for non detections; 3) "P" if the source is detected by \texttt{PySE} or "." for non detections;
    \item[source extension] 4 columns collecting the minimum and maximum of RA and dec [in deg];
    \item [Herschel-ATLAS source] counterpart name, if available within 25\,arcsec, '.' elsewhere.
\end{description}

Of the 2294 sources detected by \texttt{BLOBCAT} 1889 have a counterpart in AEGEAN, 2005 in PySE, 1678 (73.1\%) have been detected by all the three methods and only 68 (2.9\%) have been detected only by BLOBCAT.  We also released the catalog\footnote{\url{COMPONENT_CAT_LINK}} of the \texttt{BLOBCAT} $\rm{SNR}\gtrsim 3$ detections, with an extract of the outcome for all the three detection methods, when available.
For sake of reproducibility of our results, all the catalogs are released as additional material that comes together with the current paper.

\section{Radio properties and source counts}\label{sec:SourceCounts}

\begin{figure}
    \centering
    \includegraphics[width=0.45\textwidth]{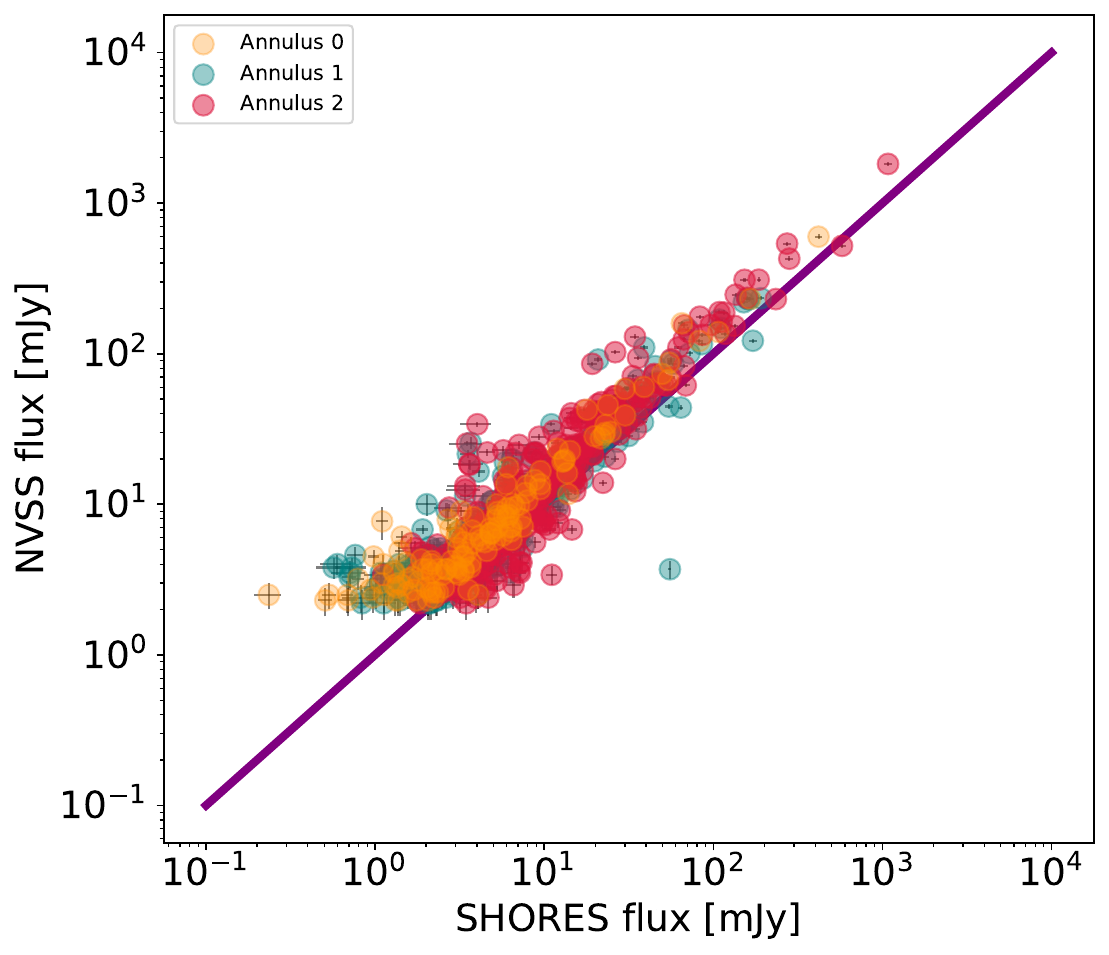}
    \includegraphics[width=0.45\textwidth]{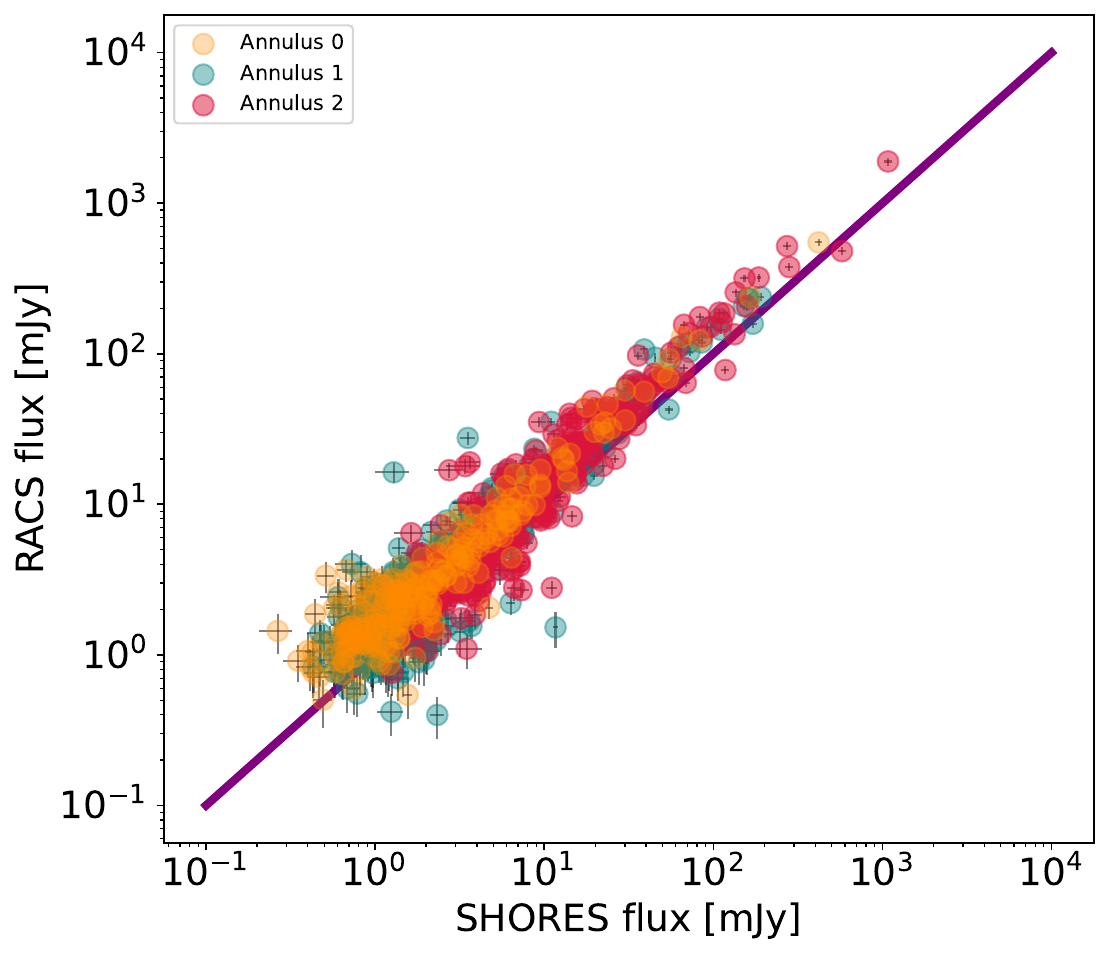}
    \caption{Fluxes extracted in the SHORES survey vs fluxes from the NVSS (\textit{left panel}) and RACS (\textit{right panel}) catalog. The different colors indicate the annulus where the source is located.}
    \label{fig:radio-ccc}
\end{figure}

\begin{figure}
    \centering
    \includegraphics[width=0.3\linewidth]{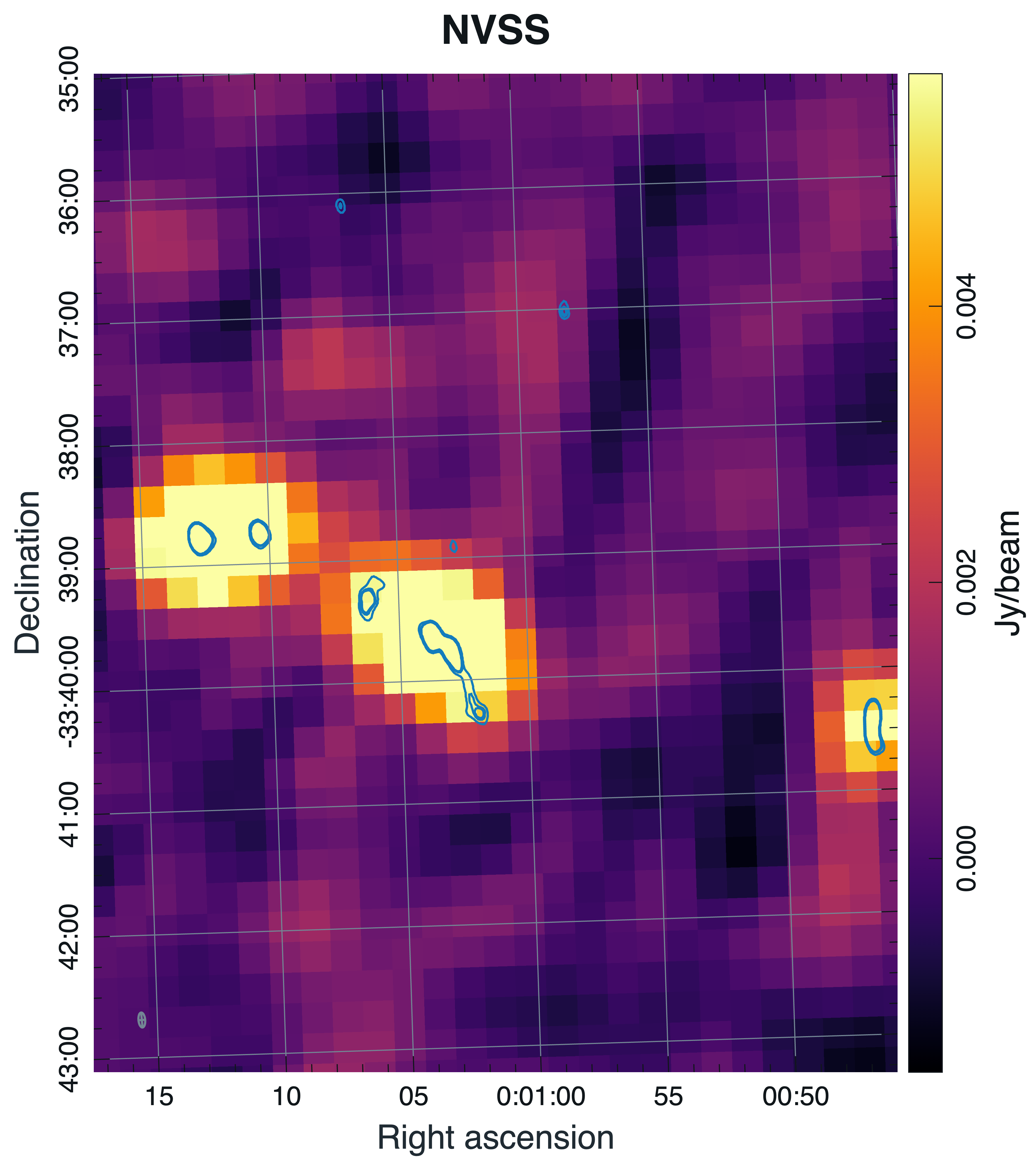}
    \includegraphics[width=0.3\linewidth]{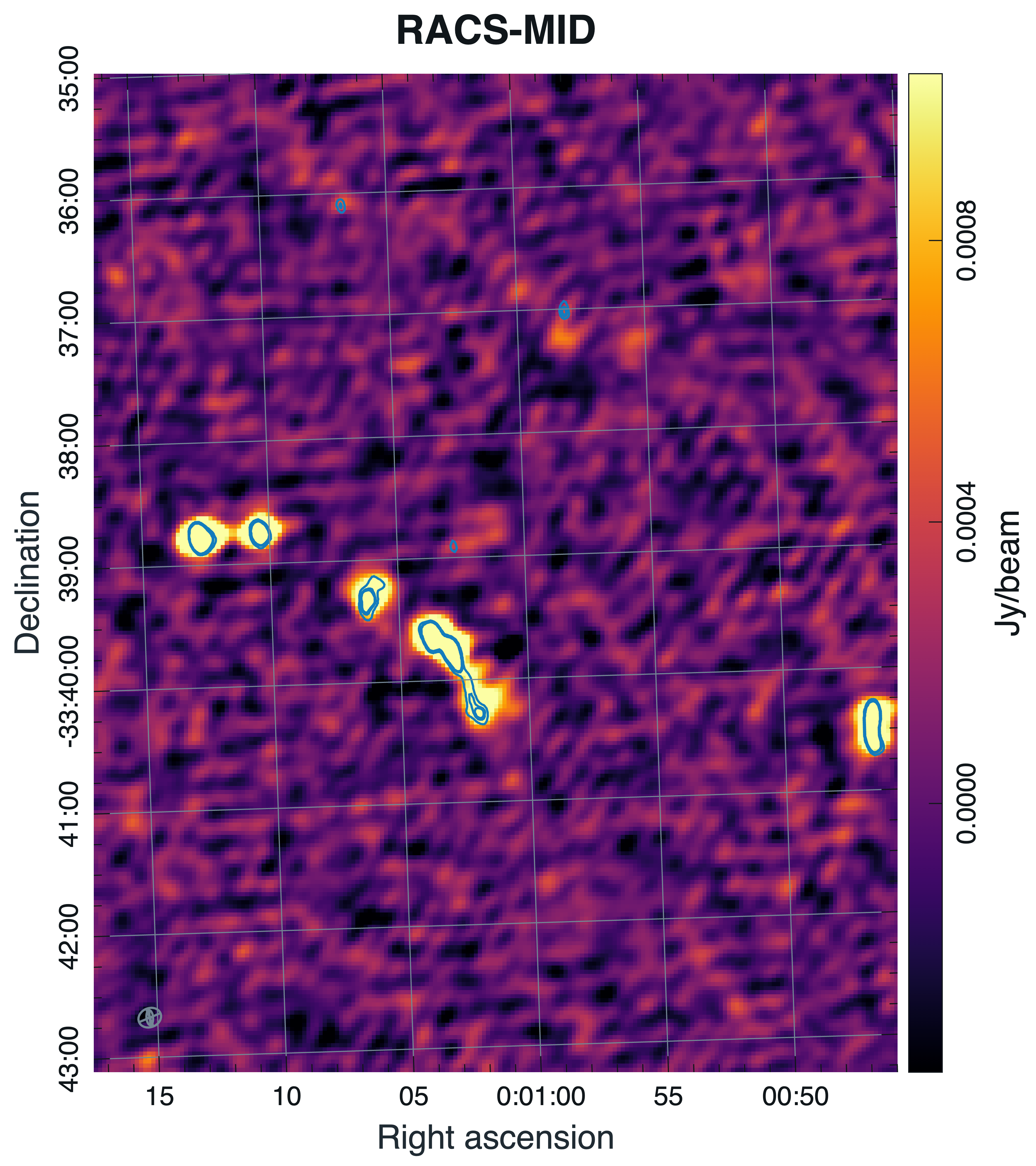}
    \includegraphics[width=0.3\linewidth]{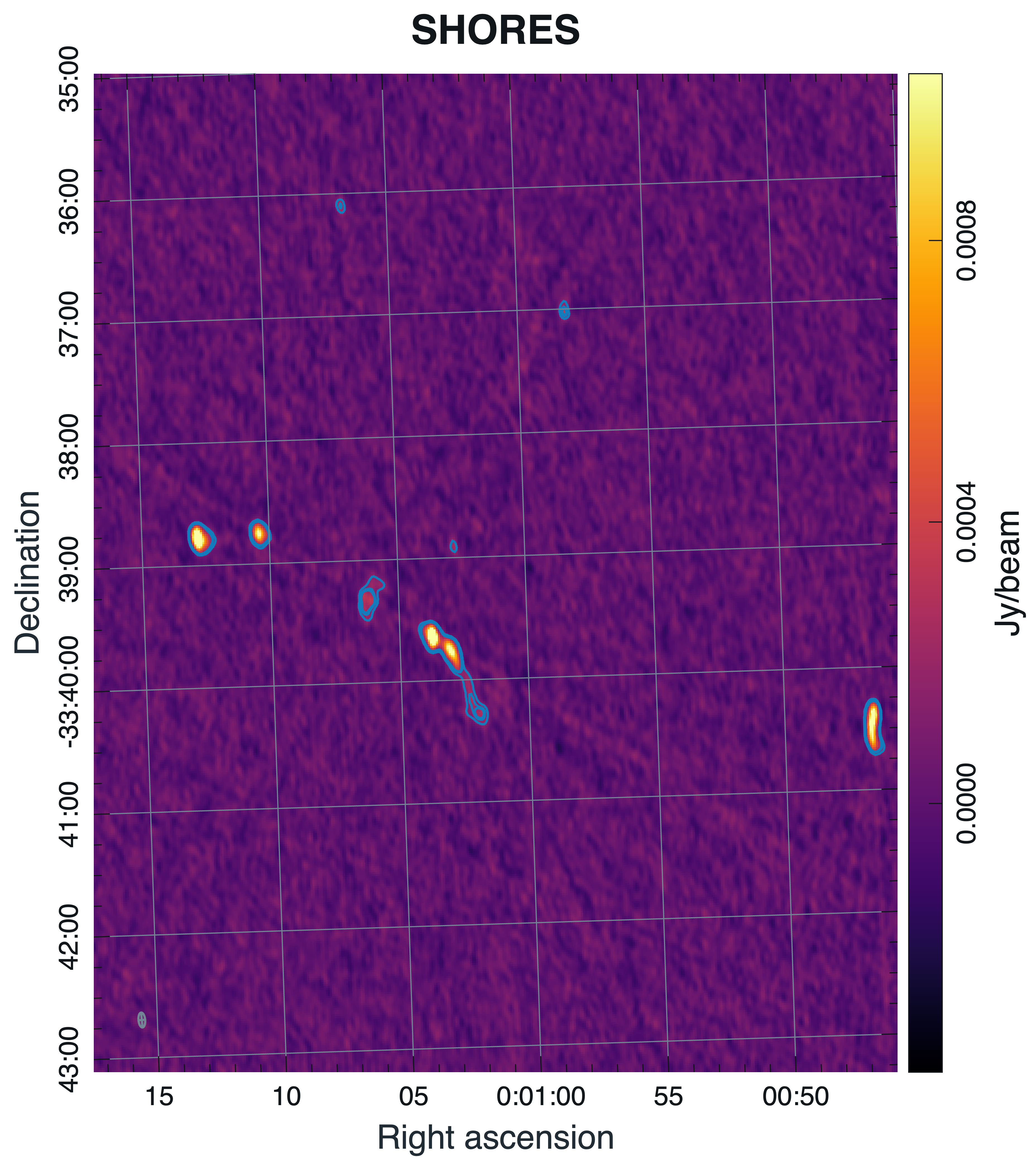}
    \caption{Comparison of the same region of field s0000-3340 as imaged by NVSS (\textit{left panel}), RACS (\textit{central panel}), and SHORES (\textit{right panel}).}
    \label{fig:nvss_comparison}
\end{figure}

As a further confirmation of our flux densities, we compared the fluxes measured by SHORES via the three different photometric tools with those reported in the literature on the same area. The NVSS 1.4 GHz survey performed with the VLA interferometer by \cite{condon98} has a FWHM resolution of 45 arcsec and a sensitivity of $2.5$ mJy/beam, cataloging $\sim 2\times 10^6$ sources. It covers $80\%$ of the celestial sphere above dec $\sim -40$.
We identified 894 sources that match within the NVSS beam size. The comparison of flux densities is in Figure \ref{fig:radio-ccc}. Correcting with a standard $-0.7$ spectral slope to account for the different frequencies of the surveys (SHORES is centered at 2.1 and partially overlaps with the NVSS 50 MHz-wide band centered at 1.4 GHz) does not fully resolve the discrepancy. However, given the sensitivity limit of NVSS around $2.5\,$mJy the common sources are mainly AGN-dominated, thus some discrepancies are possibly related to the source variability over more than $24$ years. Finally, the two surveys have a factor of $\sim 9$ in resolution ratio (our median FWHM resolution is $3.2\times 7.2$ arcsec, see the resolution comparison in Fig. \ref{fig:nvss_comparison}), and therefore NVSS could be possibly affected by blending while SHORES can be less sensitive to flux over large angular scales.

RACS is the Rapid ASKAP Continuum Survey (RACS, \citealt{mcconnell20}), a radio survey recently run with the Australian Square Kilometer Array Pathfinder (ASKAP). RACS covers the whole sky at $-90<\delta<49$ at $\sim$887.5/943.5 (RACS-low, \citealt{hale21}), 1367.5 (RACS-mid, \citealt{duchesne24}) and 1655.5 MHz (RACS-high), reaching a sensitivity of $\sim 0.25$ (RACS-low) and 0.2 mJy/beam (RACS-mid, high) with a resolution of $\sim$ 15'' (RACS-low), 10'' (RACS-mid) and 8'' (RACS-high).  Unfortunately, to the date of writing, the RACS-high catalog is not public yet, thus we compared our results with RACS-mid. Using a search radius of 8 arcsec, we found 1359 counterparts.
Figure \ref{fig:radio-ccc} shows an appreciable consistency between the fluxes extracted in this work and the ones retrieved by \cite{duchesne24}. 

\begin{table}
\centering
\begin{tabular}{l|l|l|l}
\hline
Intervals & Bin Centers & $N$ & $S_{\rm 2.1GHz}^{2.5}dN/dS$\\
mJy & mJy & & Jy$^{1.5}$sr$^{-1}$  \\
\hline
0.150 - 0.679 & 0.414 & 661 & 3.12 $\pm$ 0.12 \\
0.679 - 1.435 & 1.057 & 448 & 4.62 $\pm$ 0.22 \\
1.435 - 2.840 & 2.137 & 380 & 9.58 $\pm$ 0.49\\
2.840 - 4.500 & 3.670 & 242 & 17.31$\pm$ 1.11 \\
4.500 - 6.418 & 5.459 & 156 & 23.59 $\pm$ 1.89 \\
6.418 - 9.238 & 7.828 & 116 & 28.40 $\pm$ 2.64 \\
9.238 - 13.572 & 11.405 & 77 & 31.41 $\pm$ 3.58 \\
13.572 - 20.421 & 16.996 & 76 & 53.00 $\pm$ 6.08 \\
20.421 - 36.840 & 28.631 & 71 & 75.75 $\pm$ 8.99 \\
36.840 - 200.000 & 118.420 & 63 & 234.91 $\pm$ 29.60 \\
200.000 - 1080.000 & 640.000 & 6 & 281.20 $\pm$ 114.80 \\
\hline
\end{tabular}
\caption{Differential Euclidean source counts.}
\end{table}

\begin{figure}
    \centering
    \includegraphics[width=0.5\textwidth]{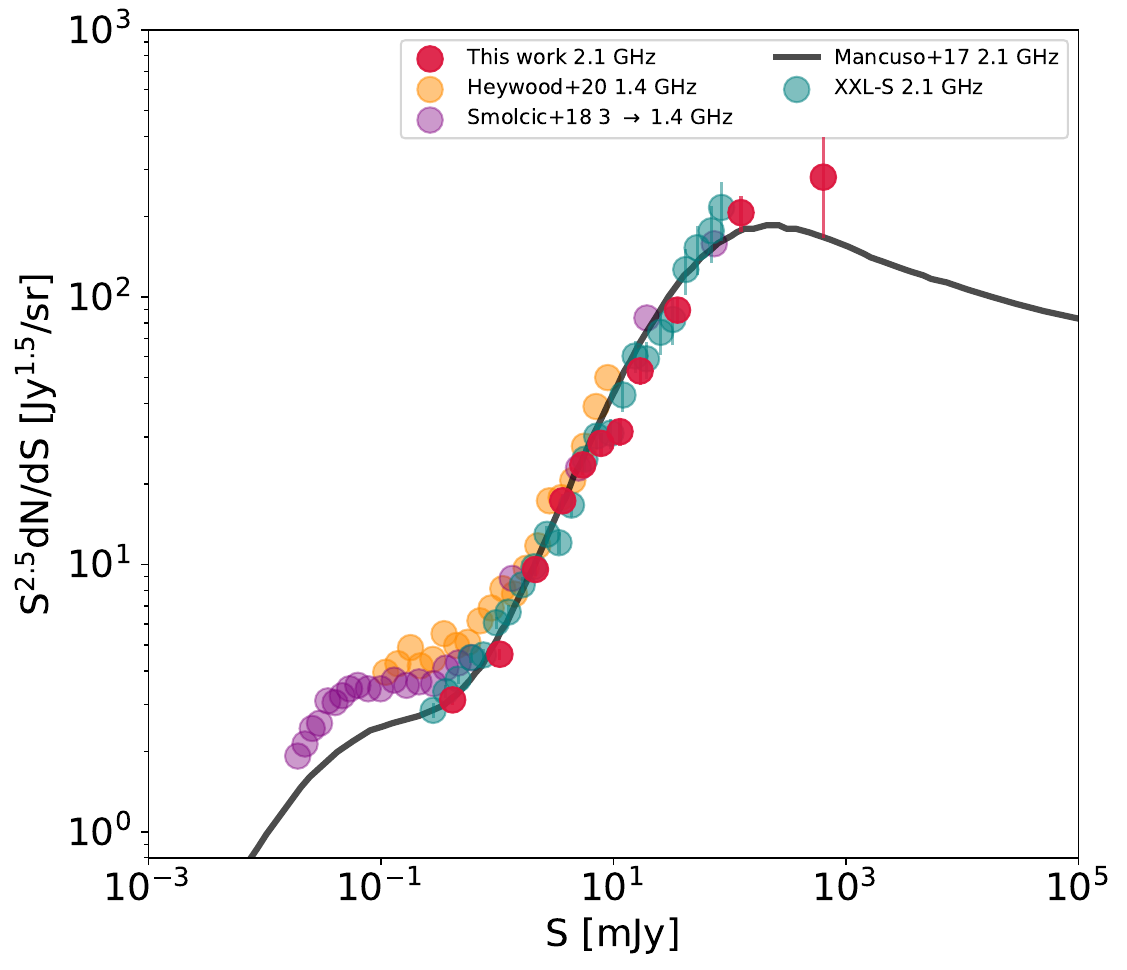}
    \caption{Euclidean differential source counts at 2.1\,GHz. For comparison we added recent estimates by \cite{butler18}, \cite{heywood20}, and \cite{smolcic18}.}
    \label{fig:counts}
\end{figure}

The Euclidean differential source counts (i.e. the number of sources per bin of flux and unit of sky area normalized to Euclidean cosmology) for the 2297 SHORES sources in the shallow fields are reported in Figure \ref{fig:counts}. The number of sources per bin of flux have been corrected for the completeness (see Figure \ref{fig:completeness}) and divided by the effective area of each of the flux bins (see Figure \ref{fig:tot_area}).
For comparison and reference, we reported the predictions by the \cite{mancuso17} model at 2.1\ GHz and past literature findings at the same frequency. 

On the one hand, our multiple independent pencil beam strategy reduces the cosmological variance effects by a factor $\sqrt{27}$ with respect to any single beam survey, and by a factor of at least 3 with respect to mosaic region surveys of comparable area (as a typically Nyquist sampled mosaic is generated by the overlap of 3 beams). On the other hand, the bright (`radio loud') population suffers from small number statistics on flux densities brighter than 100 mJy. The median flux density of our sources is 1.53 mJy. Only 25, corresponding to 1\%, of the sources are brighter than 100 mJy and the brightest source is SHORESJ234145.9-350620.5 with $1.05\pm0.15$ Jy flux density. The sub-mJy counts will be improved and considered properly in paper II dealing with the SHORES deep fields.

\section{FIR counterparts}\label{sec:FIRRC}

Unlike what happens in other bands, dust extinction does not affect radio emission, making it an optimal tracer of star formation even in the most dust-obscured objects, which are common in the early stages of galaxy formation. Star formation can produce radio emission through two main mechanisms \citep{condon92}: free–free emission from HII regions and synchrotron emission by supernova remnants.  
Star formation correlates with the presence of dust as well, as the latter is produced by massive stars and surrounds star-formation regions, in the form of molecular clouds. Dust grains absorb the ultraviolet/optical light emitted by young massive stars (types A-O) and re-emit it in the FIR bands. Thus, both FIR and radio emission correlate with ongoing star formation, thereby originating a FIR-radio correlation (FIRCC). Whenever additional processes, mainly AGN activity, trigger additional radio emission, a \textit{radio excess} with respect to the FIRRC is observed.
Considering this, the fact that SHORES fields are, by construction, covered by the FIR-survey H-ATLAS \citep{eales10} makes us able to combine the radio and FIR information of our objects. 

The FIRRC is often described via the parameter $q_{\text{FIR}}$ (e.g. \citealt{yun01} \citealt{magnelli15}, \citealt{giulietti22}) defined as:

\begin{equation}\label{eq:q_ir}
q_{\rm FIR}=\log\left(\frac{L_{\rm FIR}[\hbox{W}]/3.75\times 10^{12}}{L_{1.4\,\rm GHz} [\hbox{W}\,\hbox{Hz}^{-1}]}\right).
\end{equation}

\begin{figure}
    \centering
    \includegraphics[width=0.5\linewidth]{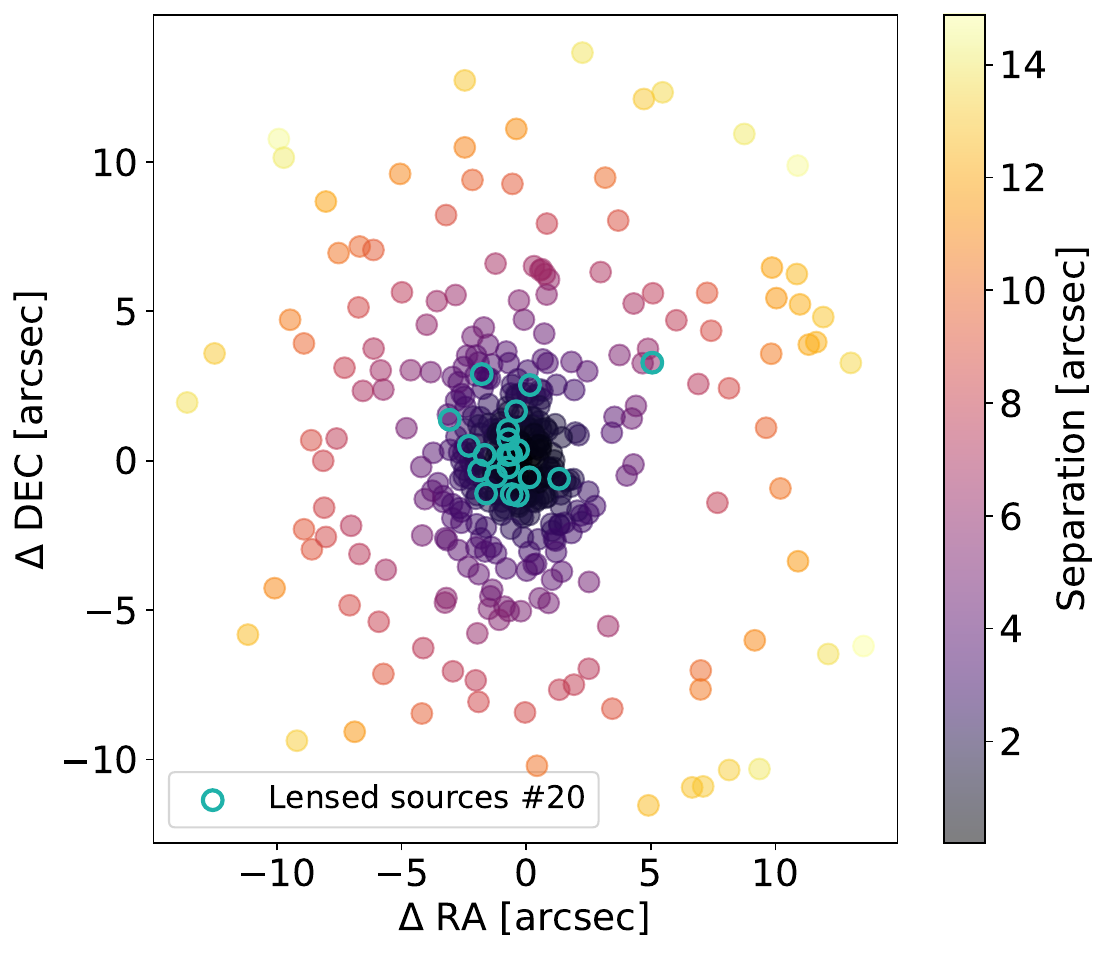}
    \caption{Discrepancy between the RA and DEC of our targets and those from the H-ATLAS catalog. The colour bar indicates the distance between the SHORES source and the corresponding H-ATLAS source.}
    \label{fig:sep_hatlas}
\end{figure}

\begin{figure}
    \centering
    \includegraphics[width=0.45\textwidth]{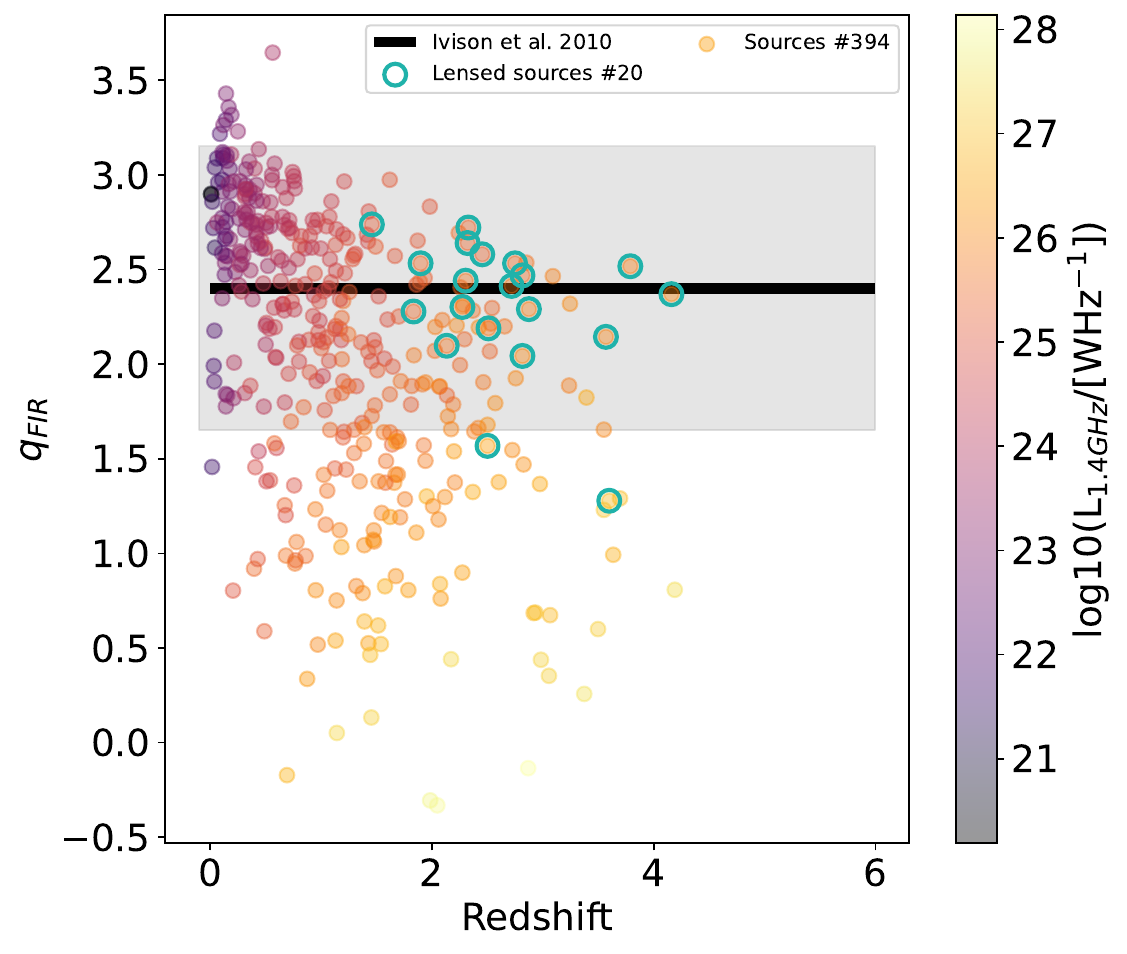}    \includegraphics[width=0.46\textwidth]{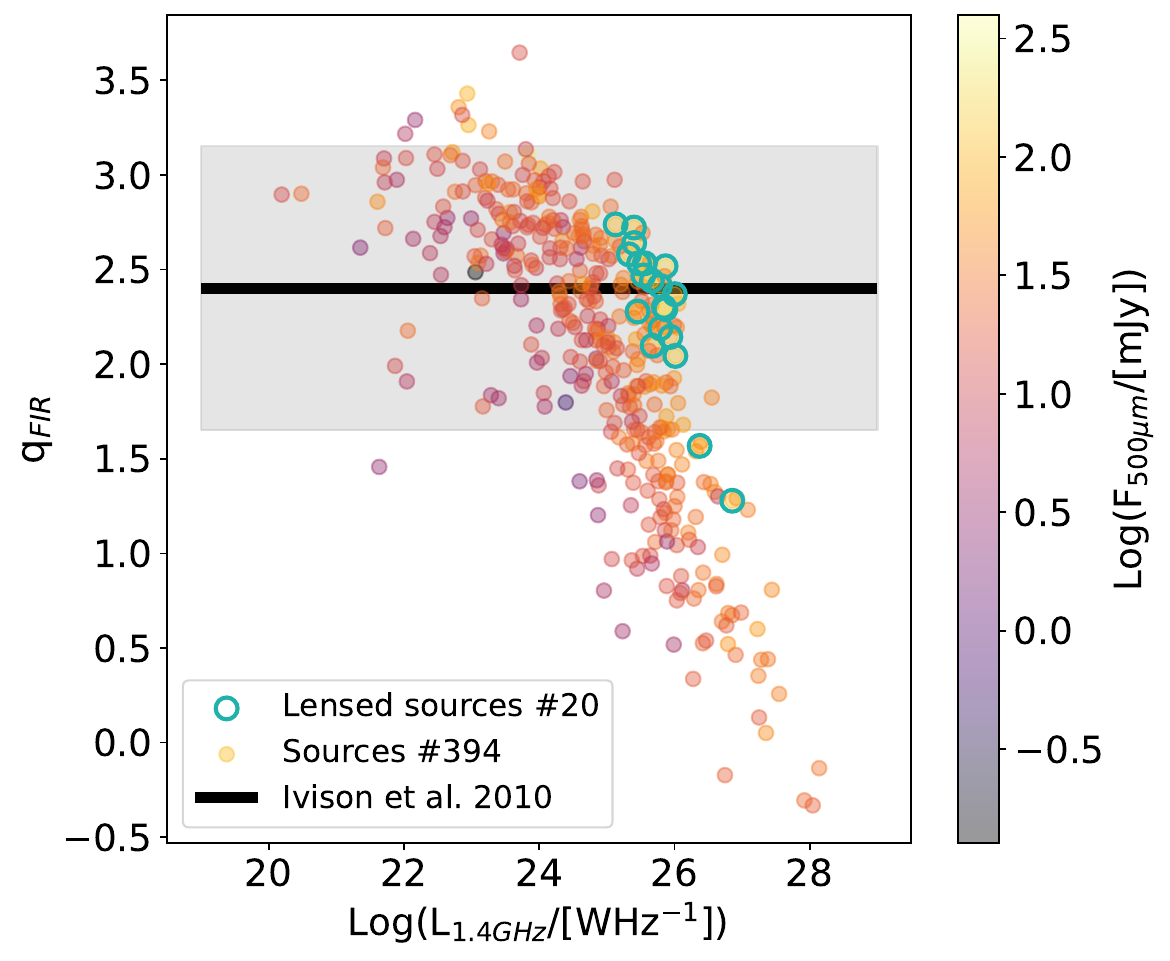}
    \caption{FIRRC $q_{FIR}$ is shown as a function of redshift (\textit{left}) and 1.4 GHz luminosity (\textit{right}) for 414 sources in our catalog, specifically those with at least three FIR photometric points, including 20 candidate lensed galaxies (cyan circles). The color bars represent the 1.4 GHz luminosity and the 500 $\mu$m flux, respectively. The fractional uncertainty on the SPIRE photometric redshift amounts to $\Delta z/(1+z)\approx 28\%$.}
    
    \label{fig:FIRRC}
\end{figure}

Therefore, we cross-matched the SHORES catalog with the H-ATLAS DR3 survey catalog, within a searching radius of 25 arcsec (the choice was motivated by a compromise between the H-ATLAS SPIRE resolution and its positional error; Fig. \ref{fig:sep_hatlas} shows that indeed our matches offset is typically below 15 arcsec). For the 457 matching sources, an estimate of the photometric redshift, based on a simple fit of the FIR data that trace the dust peak, is provided by the H-ATLAS catalog.
With that, we estimated the radio luminosity $L_{\rm 1.4 GHz}$ at 1.4 GHz for each source as
\begin{equation}
    L_{\nu,e}= \frac{4 \pi D_{L}^2(z)}{(1+z)^{1+\alpha}} \left( \frac{\nu_e}{\nu_o} \right)^{\alpha} S_{\nu, o},
\end{equation}
where $S_{\nu} \propto \nu ^{\alpha}$ is the monochromatic flux density at a certain frequency and $\alpha=-0.7$ is the typical radio slope at 1.4\,GHz for radio sources in these bands, $\nu_e$ and $\nu_o$ are the emitted and the observed frequency, and $D_L$ is the luminosity distance computed for each redshift according to the adopted $\Lambda$CDM cosmology. The FIR luminosity $L_{\rm FIR}$ is computed for each source by fitting the \textit{Herschel}/SPIRE data with a single-temperature modified black body under the optically-thin approximation with dust emissivity index $\beta =$ 1.5, the spectrum normalisation and the dust temperature (T$_{\rm dust}$) are kept as free parameters. The model ($S_{\nu, \rm best}$) which minimises the $\chi^2$ is then integrated over the wavelength range 8-1000 $\mu$m as follows: 
\begin{equation}
    L_{\rm FIR}= \frac{4\pi D_L^2}{(1+z)} \int^{1000\mu \rm m}_{8 \mu \rm m} S_{\nu,\rm best} d\nu.
\end{equation}
The resulting $q_{\rm FIR}$ as a function of redshift and radio luminosity is shown in Figure \ref{fig:FIRRC}. 
In the plots, the grey area highlights the region where star forming galaxies are preferentially located \citep{ivison10}. Nuclear activity dominated sources with an excess of radio with respect to FIR band emission appear at high radio luminosity but appreciably low $q_{\rm FIR}$.  

In the plots, we also highlight all candidate strongly lensed DSFGs with a radio counterpart, updating the results of \cite{giulietti22}. We detect radio emission for 11 additional candidate strongly lensed objects (excluding the sources in the deep fields, see Appendix \ref{sec:lensed}) which were missed in previous 2017-2019 observations, all located within $\sim 6$ arcsec to the SPIRE position. While additional high-resolution images (e.g., in the mm or optical/NIR bands) are necessary to confirm the counterpart association (see \citealt{giulietti22}), this result emphasizes the improvement of our observations' quality, which has enabled the number of lensed candidate sources with a radio counterpart to double compared to the previous sample presented in \cite{giulietti22}.



Thanks to the wealth of ancillary data available for our sources, we plan to extend this analysis by fitting their spectral energy distributions. That will allow us to test the dependence of the $q_{\rm FIR}$ from other galactic properties like the stellar mass as done by \cite{delvecchio21}. This will be particularly relevant for star-formation dominated sources in the deep fields, and so will be treated in forthcoming papers.

\section{Summary and conclusions}\label{sec:conclusions}

We introduced the Serendipitous H-ATLAS-fields Observations of Radio Extragalactic Sources (SHORES) survey that observed $29$ fields in total intensity and polarization within the \textit{Herschel}-ATLAS Southern Galactic Field via the $2$ GHz Australia Telescope Compact Aarray (ATCA) with a bandwidth centered at 2.1 GHz. 

Two of the fields have been observed to higher sensitivity. All the SHORES observations have been calibrated to account also for polarization. Polarization and deeper field analysis will be presented in future papers.  

This paper focused on the total intensity observations, calibration and analysis of the multiple pencil beam survey of the $27$ shallowest fields that cover an area of $\sim 26.1$ square degrees, with increasing sensitivity towards the phase centers of each pointing according to the ATCA 22m-dish response function, down to $\sigma \lesssim 33 \mu$Jy over a $\sim 7$ square degree region.

By fitting the spectral behavior of sources detected in each of the four 512 MHz-wide sub-bands in which we divided our observing bandwidth, we determined a refinement of the ATCA primary beam correction that allows to extract sources over a range of 3 FWHM of the FOV for each field. Given the telescope response, only the brightest sources can be detected at the largest distances, but still this improved our source counts determination by increasing the statistics of bright sources and reducing sampling variance.

We tested the consistency of our procedures by matching multiple detection tools, by mean of realistic simulations and by comparing our results with NVSS and RACS catalogs.
   
The final SHORES shallow-field sample counts 2294 sources detected with \texttt{BLOBCAT} to signal-to-noise ratio $\rm{SNR}\gtrsim 4.5$. Simulations determined that our procedure and final catalog is $95\%$ reliable to the above significance level and $95\%$ complete above $0.5$ mJy. By exploiting ATCA E-W 6km-configuration we reached resolutions of $3.2\times 7.2$ arcsec, to which level $81\%$ of our sources are unresolved. 

We determined 2.1 GHz source counts down to the $150\,\mu$Jy level. They are in good agreement with previous determinations. Thanks to the overlap with the H-ATLAS survey we could exploit FIR information on 394 matching sources to obtain a photometric redshift estimate, and luminosities in the radio and FIR regime so determining the FIR-radio Correlation. The results expands to unlensed sources what was found by \cite{giulietti22} for the lensed candidates targeted in the ATCA project that constituted the pilot of this SHORES survey.  

\begin{acknowledgments}
We thank the anonymous referee for the constructive and detailed comments, that appreciably improved our manuscript.
We acknowledge Q. D'Amato for his contributions in the early stages of the SHORES survey. We are indebted to E. Mahony, E. Sadler, T. Murphy, P. Hancock, J. Stevens, and R. Ekers for critical reading and their support to the project. 
We also thank the "Crema \& Gusto" site in Funo (Bologna, Italy) and the "Sirena" establishment in Grignano (Trieste, Italy) for logistically supporting the SHORES team. This work was partially funded from the projects: INAF GO-GTO Normal 2023 funding scheme with the project "Serendipitous H-ATLAS-fields Observations of Radio Extragalactic Sources (SHORES)";  INAF Large Grant 2022 funding scheme with the project "MeerKAT and LOFAR Team up: a Unique Radio Window on Galaxy/AGN co-Evolution; ``Data Science methods for MultiMessenger Astrophysics \& Multi-Survey Cosmology'' funded by the Italian Ministry of University and Research, Programmazione triennale 2021/2023 (DM n.2503 dd. 9 December 2019), Programma Congiunto Scuole; EU H2020-MSCA-ITN-2019 n. 860744 \textit{BiD4BESt: Big Data applications for black hole Evolution STudies}; Italian Research Center on High Performance Computing Big Data and Quantum Computing (ICSC), project funded by European Union - NextGenerationEU - and National Recovery and Resilience Plan (NRRP) - Mission 4 Component 2 within the activities of Spoke 3 (Astrophysics and Cosmos Observations).
The Australia Telescope Compact Array is part of the Australia Telescope National Facility (https://ror.org/05qajvd42) which is funded by the Australian Government for operation as a National Facility managed by CSIRO. We acknowledge the Gomeroi people as the Traditional Owners of the Observatory site. 
This scientific work uses data obtained from Inyarrimanha Ilgari Bundara / the Murchison Radio-astronomy Observatory. We acknowledge the Wajarri Yamaji People as the Traditional Owners and native title holders of the Observatory site. CSIRO’s ASKAP radio telescope is part of the Australia Telescope National Facility (https://ror.org/05qajvd42). Operation of ASKAP is funded by the Australian Government with support from the National Collaborative Research Infrastructure Strategy. ASKAP uses the resources of the Pawsey Supercomputing Research Centre. Establishment of ASKAP, Inyarrimanha Ilgari Bundara, the CSIRO Murchison Radio-astronomy Observatory and the Pawsey Supercomputing Research Centre are initiatives of the Australian Government, with support from the Government of Western Australia and the Science and Industry Endowment Fund. This paper includes archived data obtained through the CSIRO ASKAP Science Data Archive, CASDA (https://data.csiro.au).
\end{acknowledgments}

\vspace{5mm}
\facilities{ATCA}
\software{\texttt{AEGEAN} \citep{aegean},  
            \texttt{BLOBCAT} \citep{blobcat}
            \texttt{Miriad} \citep{miriad}
            \texttt{PySE} \citep{pyse}
            \texttt{WSCLEAN} \citep{offringa14}}

\bibliography{bibliography}{}
\bibliographystyle{aasjournal}



\appendix

\section{Lensed galaxies candidates at the center of the fields}\label{sec:lensed}

As an example of the properties of SHORES surveys we collected the information for the
H-ATLAS galaxies selected by \cite{negrello17} to be lensed candidate that we targeted at the center of our SHORES fields (see Figure \ref{fig:map}).

\begin{figure*}
    \centering
    \includegraphics[width=\linewidth]{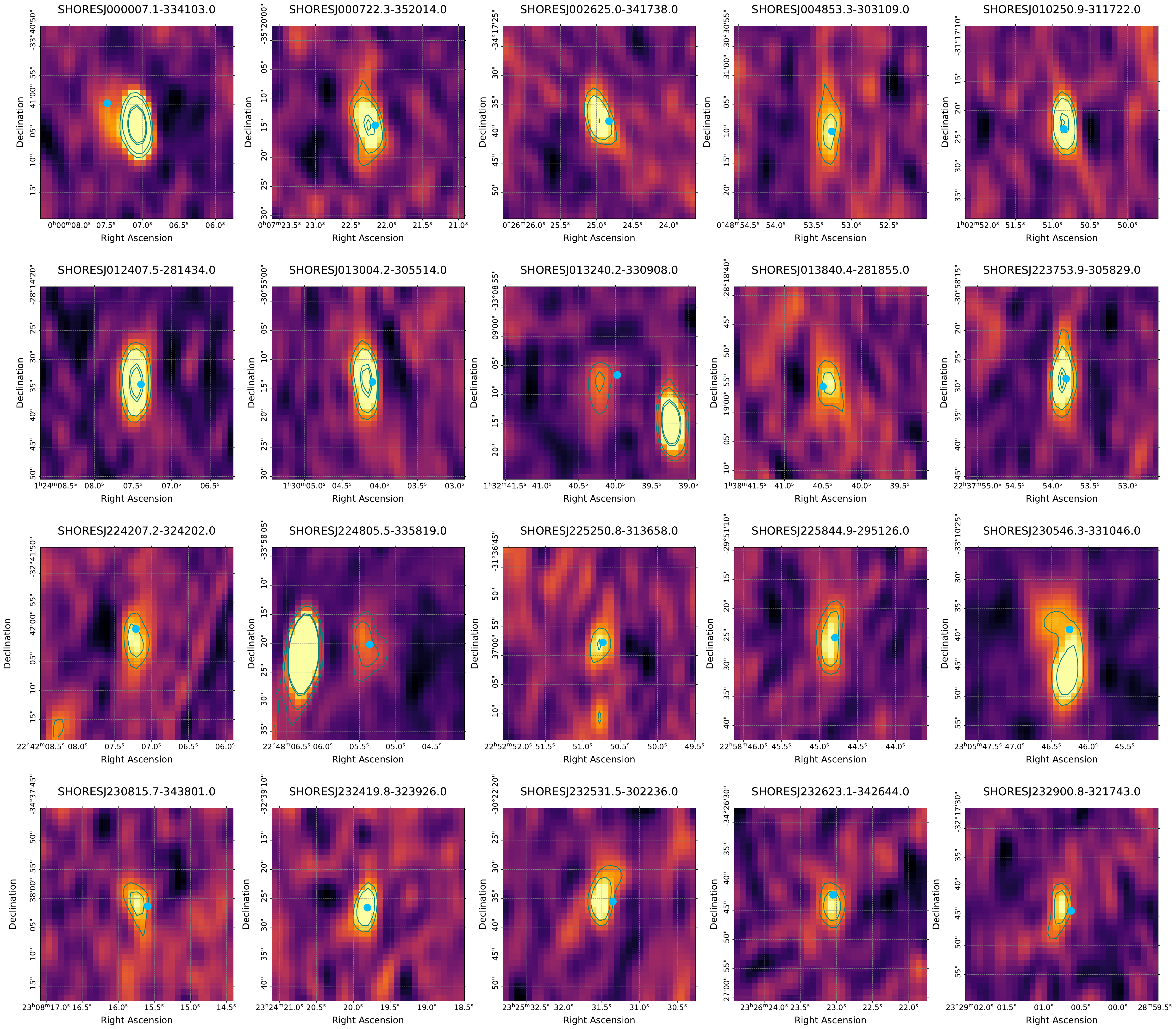}
    \caption{Snapshot of SHORES map in the position of the candidate lensed sources at the center of the SHORES fields with a detection with signal to noise ratio $>4.5$.}
    \label{fig:lensed}
\end{figure*}

Of the 27 FIR-selected lensed candidates in the center of SHORES shallow fields, 20 are detected with a SNR$>$4.5 in our survey. Compared to the data presented by \cite{giulietti22} that was based on the 2 earliest SHORES epochs (see Table \ref{tab:epochs}), we almost doubled the number of objects with an ATCA detection. We show their radio and FIR fluxes in Table
\ref{tab:lensed} and the radio maps in Figure \ref{fig:lensed}. 
As shown in Figure \ref{fig:FIRRC}, the majority of the candidated lensed sources have a $q_{FIR}$ typical of DSFGs, confirming the hypothesis that the radio emission could be mostly due to star-formation activity as already suggested by \cite{giulietti22}.

\begin{sidewaystable}
\centering
\scriptsize
\begin{tabular}{c|c|c|c|c|c|c|c}
\hline
SHORES ID & H-ATLAS ID & S$_{250\mu m}$ & S$_{350\mu m}$ & S$_{500\mu m}$ &  S$_{2.1GHz}$ & Separation & z$_{\rm SPIRE}$ \\
 & & mJy & mJy & mJy &  mJy & arcsec &  \\
\hline
SHORESJ000007.1-334103.0 & HATLASJ000007.5-334100 & 130.27$\pm$7.57 & 160.04$\pm$8.39 & 116.25$\pm$8.42 & 2.05$\pm$0.12 & 6.03 & 2.5$\pm$0.42  \\ 
SHORESJ000722.3-352014.0 & HATLASJ000722.2-352015 & 237.37$\pm$7.46 & 192.87$\pm$8.2  & 107.53$\pm$8.55 & 0.5$\pm$0.05  & 1.31 & 1.46$\pm$0.3  \\ 
SHORESJ002625.0-341738.0 & HATLASJ002624.8-341738 & 137.69$\pm$7.46 & 185.94$\pm$8.34 & 148.81$\pm$8.7  & 0.44$\pm$0.05 & 1.67 & 2.87$\pm$0.46 \\ 
SHORESJ004853.3-303109.0 & HATLASJ004853.3-303110 & 118.08$\pm$6.94 & 147.33$\pm$7.76 & 105.44$\pm$8.05 & 0.53$\pm$0.05 & 1.26 & 2.51$\pm$0.42 \\ 
SHORESJ010250.9-311722.0 & HATLASJ010250.9-311723 & 267.87$\pm$7.45 & 253.15$\pm$8.26 & 168.07$\pm$8.93 & 0.55$\pm$0.05 & 1.22 & 1.9$\pm$0.35  \\ 
SHORESJ012407.5-281434.0 & HATLASJ012407.4-281434 & 257.49$\pm$8.07 & 271.1$\pm$8.53  & 203.96$\pm$8.68 & 0.78$\pm$0.06 & 0.75 & 2.28$\pm$0.39 \\ 
SHORESJ013004.2-305514.0 & HATLASJ013004.1-305514 & 164.4 $\pm$6.8  & 147.52$\pm$7.87 & 100.6$\pm$8.01  & 0.57$\pm$0.05 & 0.76 & 1.83$\pm$0.34 \\ 
SHORESJ013240.2-330908.0 & HATLASJ013240.0-330907 & 112.02$\pm$7.64 & 148.82$\pm$8.68 & 117.74$\pm$8.83 & 0.22$\pm$0.04 & 3.35 & 2.82$\pm$0.46 \\ 
SHORESJ013840.4-281855.0 & HATLASJ013840.5-281856 & 116.28$\pm$7.85 & 177.03$\pm$8.54 & 179.34$\pm$8.96 & 0.21$\pm$0.04 & 1.45 & 3.79$\pm$0.57 \\ 
SHORESJ223753.9-305829.0 & HATLASJ223753.8-305828 & 139.07$\pm$7.24 & 144.77$\pm$7.87 & 100.55$\pm$8.26 & 0.64$\pm$0.06 & 1.01 & 2.13$\pm$0.38 \\ 
SHORESJ224207.2-324202.0 & HATLASJ224207.2-324159 & 72.96 $\pm$7.67 & 88.14$\pm$8.64  & 100.77$\pm$9.42 & 0.29$\pm$0.05 & 2.54 & 3.57$\pm$0.55 \\ 
SHORESJ224805.5-335819.0 & HATLASJ224805.4-335820 & 122.33$\pm$7.82 & 135.56$\pm$8.74 & 126.92$\pm$9.0  & 0.65$\pm$0.05 & 0.47 & 2.82$\pm$0.46 \\ 
SHORESJ225250.8-313658.0 & HATLASJ225250.7-313658 & 127.35$\pm$6.73 & 138.74$\pm$7.7  & 111.4$\pm$8.0   & 0.19$\pm$0.04 & 0.64 & 2.45$\pm$0.41 \\ 
SHORESJ225844.9-295126.0 & HATLASJ225844.8-295125 & 175.39$\pm$7.43 & 186.94$\pm$8.42 & 142.63$\pm$9.26 & 0.27$\pm$0.04 & 1.26 & 2.32$\pm$0.4  \\ 
SHORESJ230546.3-331046.0 & HATLASJ230546.3-331039 & 76.78$\pm$7.73  & 110.9$\pm$8.43  & 110.43$\pm$8.82 & 2.33$\pm$0.12 & 3.4  & 3.6$\pm$0.55  \\ 
SHORESJ230815.7-343801.0 & HATLASJ230815.6-343801 & 79.37 $\pm$7.62 & 135.4$\pm$8.28  & 139.96$\pm$8.87 & 0.22$\pm$0.04 & 1.91 & 4.16$\pm$0.62 \\ 
SHORESJ232419.8-323926.0 & HATLASJ232419.8-323927 & 212.93$\pm$6.83 & 244.18$\pm$7.62 & 169.37$\pm$7.92 & 0.5$\pm$0.06  & 0.57 & 2.3$\pm$0.4  \\ 
SHORESJ232531.5-302236.0 & HATLASJ232531.4-302236 & 175.55$\pm$6.8  & 227.0$\pm$7.6   & 175.72$\pm$7.87 & 0.42$\pm$0.05 & 2.35 & 2.72$\pm$0.45 \\ 
SHORESJ232623.1-342644.0 & HATLASJ232623.0-342642 & 153.7 $\pm$6.87 & 178.35$\pm$7.77 & 123.48$\pm$8.2  & 0.27$\pm$0.04 & 1.7  & 2.33$\pm$0.4  \\ 
SHORESJ232900.8-321743.0 & HATLASJ232900.6-321744 & 118.31$\pm$7.05 & 141.25$\pm$7.94 & 119.69$\pm$8.39 & 0.24$\pm$0.04 & 1.94 & 2.75$\pm$0.45 \\ 
\hline
\end{tabular}
\caption{SHORES and \textit{Herschel}-ATLAS information for the candidate lensed galaxies at the center of SHORES fields. 
}
\label{tab:lensed}
\end{sidewaystable}

\end{document}